\documentclass[%
reprint,
 amsmath,amssymb,
 aps,
floatfix,
]{revtex4-2}

\usepackage[english]{babel}
\usepackage{comment}
\usepackage[english]{babel}
\makeatletter
\@namedef{l@en}{\l@english}
\makeatother
\usepackage[toc,page]{appendix}

\usepackage{amsmath}
\usepackage{amssymb}

\usepackage{cleveref}
\usepackage{graphicx}
\usepackage{dcolumn}
\usepackage{bm}
\usepackage{xcolor}
\usepackage[dvipsnames]{xcolor}

\newcommand{\bs}[1]{\boldsymbol{#1}}

\begin{document}

\title{Noise enhances odor source localization}

\author{Francesco Marcolli}
\author{Martin James}%
\author{Agnese Seminara}%
 \email{agnese.seminara@unige.it}
\affiliation{Machine Learning Genoa Center (MaLGa) \& Department of Civil, Chemical and Environmental Engineering, University of Genoa, Genoa, Italy}

\date{January 12, 2026}

\begin{abstract}

We address the problem of inferring the location of a target that releases odor in the presence of turbulence. Input for the inference is provided by many sensors scattered within the odor plume. Drawing inspiration from distributed chemosensation in biology, we ask whether the accuracy of the inference is affected by proprioceptive noise, i.e., noise on the perceived location of the sensors.
Surprisingly, in the presence of a net fluid flow, proprioceptive noise improves Bayesian inference, rather than degrading it. An optimal noise exists that efficiently leverages additional information hidden within the geometry of the odor plume. 
Empirical tuning of noise functions well across a range of distances and may be implemented in practice. 
Other sources of noise also improve accuracy, owing to their ability to break the spatiotemporal correlations of the turbulent plume. These counterintuitive benefits of noise may be leveraged to improve sensory processing in biology and robotics. 
\end{abstract}

\maketitle

\section{Introduction}

Octopuses, like other cephalopods, are equipped with distributed sensory systems: their eight arms are covered by suckers that contain thousands of sensory cells detecting chemical and mechanical cues both from surfaces and from the water itself~\cite{van2020molecular, maselli2020sensorial, allard2023structural, kang2023sensory, villanueva2017cephalopods, al2021identification,tarvin2020sucker,weertman2024octopus, sivitilli2023mechanisms}. 
Chemical information facilitates prey location in the laboratory
even in the absence of vision~\cite{weertman2024octopus}. 
In these experiments, chemicals released by the prey are carried by water in a noisy and sparse plume, akin to realistic conditions in the field that are often dominated by turbulence. 
Thus, these organisms use distributed chemosensation as a distal sense for turbulent navigation. To this end, information must be extracted from an array of chemosensory cells scattered throughout their body. A similar problem is widely studied in robotics, where distributed information from an array of chemical sensors is used to e.g.~localize chemical leaks~\cite{russell1995robotic,kowadlo2008robot,chen2019odor}. \\

\noindent 
To process distributed information, algorithms for robotics usually combine sensors' individual readouts with their location in space~\cite{nehorai1995detection,alpay2002model,matthes2005source}. 
Likewise, animals often combine the individual signal from each sensory cell with knowledge of the cell's location in space, encoded within topographic maps of the sensory organ in the brain~\cite{hubel1962receptive, hubel1959receptive,  decharms1996primary, romani1982tonotopic, woolsey1970structural, petersen2007functional, petersen2012attention}.

\noindent In contrast, octopus appears to lack somatotopic maps of sensory input and motor control in the brain and this has been linked to the challenges of encoding the shape of a flexible body~\cite{zullo2009nonsomatotopic,tytell}. 
Indeed, the position of a cell within a sucker is not sufficient to localize the cell in space, as the sucker is free to move substantially relative to the rest of the body. A map of the organ itself would need to be coupled to proprioception, the sense that enables an animal to perceive the location of its body parts in space. Despite the lack of somatotopy, behavioral experiments suggest the octopus is likely equipped with some degree of proprioception~\cite{gutfreund1996,gutfreund1998,gutnick2020use,zullo2011new, hochner2013nervous}, potentially facilitated by an embodied mechanism~\cite{hochner2023embodied}. However, to what extent precise proprioceptive feedback is transmitted to the brain 
remains to be established~\cite{zullo2025}.
Thus, octopuses extract useful information from distributed measures of turbulent odors, despite potentially noisy information on sensors' position in space. But is the positional information of many chemosensors crucial to interpret their collective readout? Answering this question will be directly relevant for soft robotics, to inform the design of octopus-inspired algorithms for multi-sensor localization of chemical sources.
Inspired by distributed chemosensation in octopus, we consider a virtual agent composed of many sensors that detect smell in the presence of turbulence, obtained from Direct Numerical Simulations. We ask whether accurate information regarding the position of the sensors is essential to infer the location of the olfactory target that releases the odor. Surprisingly, positional noise does not deteriorate instantaneous Bayesian inference of target location, but rather improves its accuracy. 
We show that noisy inference leverages additional information hidden in the anisotropy of the odor in the presence of a wind or current. An optimal level of noise exists, which depends on the unknown target location.
By developing an asymptotic theory, we show that noise can be tuned empirically to approximate the unknown optimal noise. Alternative sources of positional and sensory noise also improve inference and function by partially curing misspecification of the model. \\ 
Our results imply that accurate proprioception is not always essential to process distributed chemosensation in the presence of turbulence. On the contrary, living and non-living systems may exploit noise to improve the predictive power of their sensory systems. 

\begin{figure}[h!]
\includegraphics[width=\linewidth]{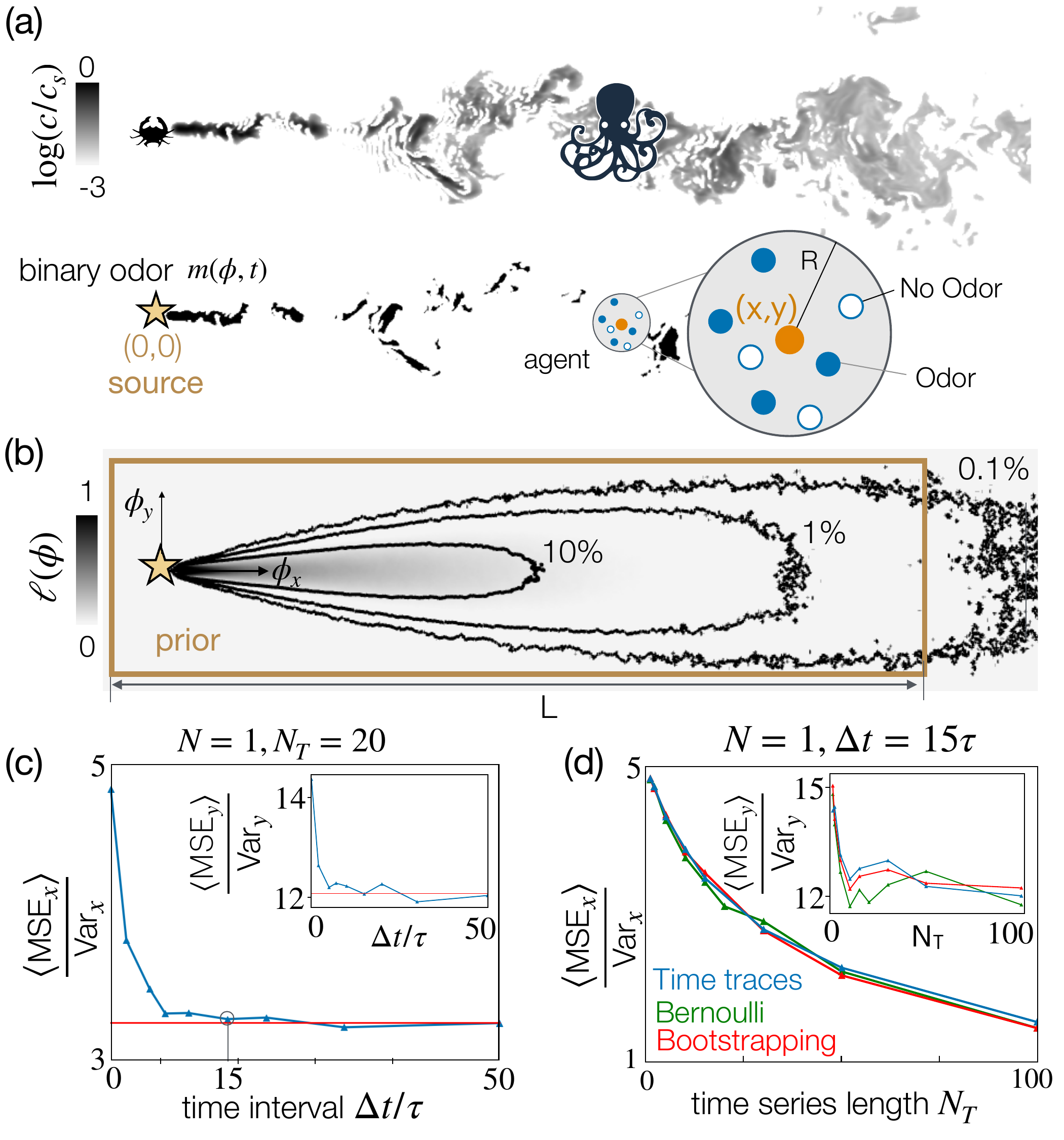} 
\caption{\label{fig:setting} 
(a) A snapshot of the odor field $c(\bs\phi)$ emitted from an olfactory target (crab), obtained through direct numerical simulations (top) and its binarized version $m(\bs\phi)$ (bottom). The multisensor agent is represented as a circle of radius $R$ centered in $(x,y)$;  full and empty dots represent the $N$ sensors, either detecting or non-detecting odor. Our goal is to infer the coordinates $(x,y)$ of the center from binary odor detections. (b) Empirical likelihood $\ell(\bs\phi)=\langle m(\bs\phi)\rangle$. The brown rectangle delimits a uniform prior. (c) Mean Squared Error in the $x$ coordinate (main plot) and in the $y$ coordinate (inset), averaged over all 52 points $(x,y)$: $\langle \text{MSE}_x\rangle = 1/52 \sum_{i=1}^{52} \text{MSE}_x^i $, for inference with a single sensor and $N_T=20$ time points, sampled at regular intervals $\Delta t$ (blue line). 
For $\Delta t \gtrsim 15 \tau$, $\langle {\text{MSE}_x} \rangle$ plateaus to $3.3 \text{Var}_x$, matching predictive accuracy with the same number of time points sampled randomly in the whole simulation (red line), confirming measures are uncorrelated. (d) Same as (c) for a single sensor and an increasing number of time points $N_T$ sampled at an equal interval $\Delta t = 15 \tau $.} 
\end{figure}

\section{Results}
\label{sec:results}

\noindent \textbf{Modeling and methodology:}
We consider an olfactory agent composed of a collection of $N$ sensors randomly distributed inside a circle of radius $R$ centered at $\bs\xi=(x,y)$. Sensor $i$ is located at  $\bs{\xi}_i=\bs\xi + (x_i,y_i)$, where $(x_i,y_i)$ are coordinates relative to the center. At the origin of the coordinate system, there is a target that releases odor in the presence of turbulence (Fig.~\ref{fig:setting}a). The odor field $c(\bs{\phi},t)$ at any time $t$ and location $\bs{\phi}$  is obtained from computational fluid dynamics simulations of a turbulent open channel flow, the details of which are provided in the appendix. In the entire manuscript, spatial scales are provided in units of $\Delta x$, the grid spacing for the simulations (see Table~\ref{tab:simulation_parameters}). The agent has a sensitivity threshold $c_0$, and performs binary measurements, $m$, (Fig.~\ref{fig:setting}a, bottom) according to: 
\begin{equation}
 m(\bs \phi,t)=   
\begin{cases} 
       1, & \text{if} \; \; c(\bs \phi,t) \ge c_0\ \ \text{and}\\
       0 &  \text{otherwise.}
     \end{cases}
     \label{eq:thresholding}
\end{equation}
We assume that the agent has an empirical model of the likelihood to detect odor at any location $\bs\phi$:
\begin{equation}
    \ell(\bs\phi)=p \left( m=1\:| \: \bs \phi \right) 
    =\bar{ m}(\bs \phi)
    \label{eq:probability_map}
\end{equation}
where $\bar{m}(\bs\phi)$ denotes the temporal average of $m$ and the odor field $c$ is assumed to be statistically stationary (Fig.~\ref{fig:setting}b). We will compute  $\bar{m}(\bs\phi)$ as an empirical average of two-dimensional odor snapshots from numerical simulations, except for the theoretical results where we use Eq.~\eqref{eq:power_law_map}.

\noindent Using Bayesian inference from $N$ independent sensors~\cite{KEATS2007465}, the agent infers the position of its center to be at $\hat{\bs\xi} = \arg \max_{\bs \xi} \prod_{i=1}^N L(m_i|\bs{\xi}) p(\bs{\xi})$, where $p(\bs{\xi})$ is a uniform prior (brown rectangle in Fig.~\ref{fig:setting}b). Inference is in two dimensions, as $\hat{\bs\xi}= (\hat{x},\hat{y})$. Here, $L(m_i|\bs{\xi})$ is the likelihood that the $i-$th sensor detects $m_i$ given the center is at $\bs\xi$:
\begin{equation}\label{eq:L}
    L(m_{i}|\bs{\xi}) =\int \ell_{i}^{m_{i}} (1-\ell_{i})^{1-m_{i}}p(\bs{\phi}_i|\bs{\xi}) d\bs{\phi}_i
\end{equation}
\noindent
where $\ell_i=\ell(\bs\phi_i)$ depends on physics and $p(\bs{\phi}_i|\bs{\xi})$ models positional information, or proprioception. We start by modelling two degrees of positional information: 
\begin{equation}\label{prop_cases_eq}
p(\bs{\phi}_i|\bs{\xi}) = 
    \begin{cases} 
            \delta (\bs{\phi}_i - \bs{\xi}_i(\bs{\xi})) \: \: \: \text{Perfect}\\
       \delta (\bs{\phi}_i - \tilde{\bs{\xi}}_i(\bs{\xi})) \: \: \: \text{Noisy}   
    
        \end{cases}        
\end{equation}
\noindent 
where $\tilde{\bs{\xi}}_i$ are perceived sensor positions relative to the center, corresponding to noisy measures of the real relative positions: $\tilde{\bs{\xi}}_i = \bs{\xi}_i + \bs{\gamma}_i$. Here $\bs{\gamma}_i\sim  \mathcal{N}_T(0,\eta^2)$ are drawn from a Gaussian with zero mean and standard deviation $\eta$, truncated between $\left(-\eta-1, \eta+1\right)$. 
Thus, noisy proprioception depicts a scenario where the agent perceives its sensors to be at locations $\tilde{\bs\xi}_i$, and is not aware that these are not the correct locations, ${\bs\xi}_i$. In contrast, perfect proprioception corresponds to the case where the agent leverages precise information of the relative positions between its sensors.\\ 
To quantify the accuracy of the inference, we evaluate the mean squared error MSE$_x=\frac{1}{N_r}\sum_{i=1}^{N_r} (\hat x_i - x)^2$ and bias $b_x=\frac{1}{N_r}\sum_{i=1}^{N_r} (\hat x_i - x )$ between the inferred position and the real position (and similarly for the $y$ coordinate) --  here $i$ runs over the $N_r$ realizations of the inference for the same location $\bs\xi =(x,y)$ of the target. \\ 

\begin{figure}[h!]
\includegraphics[width=\linewidth]{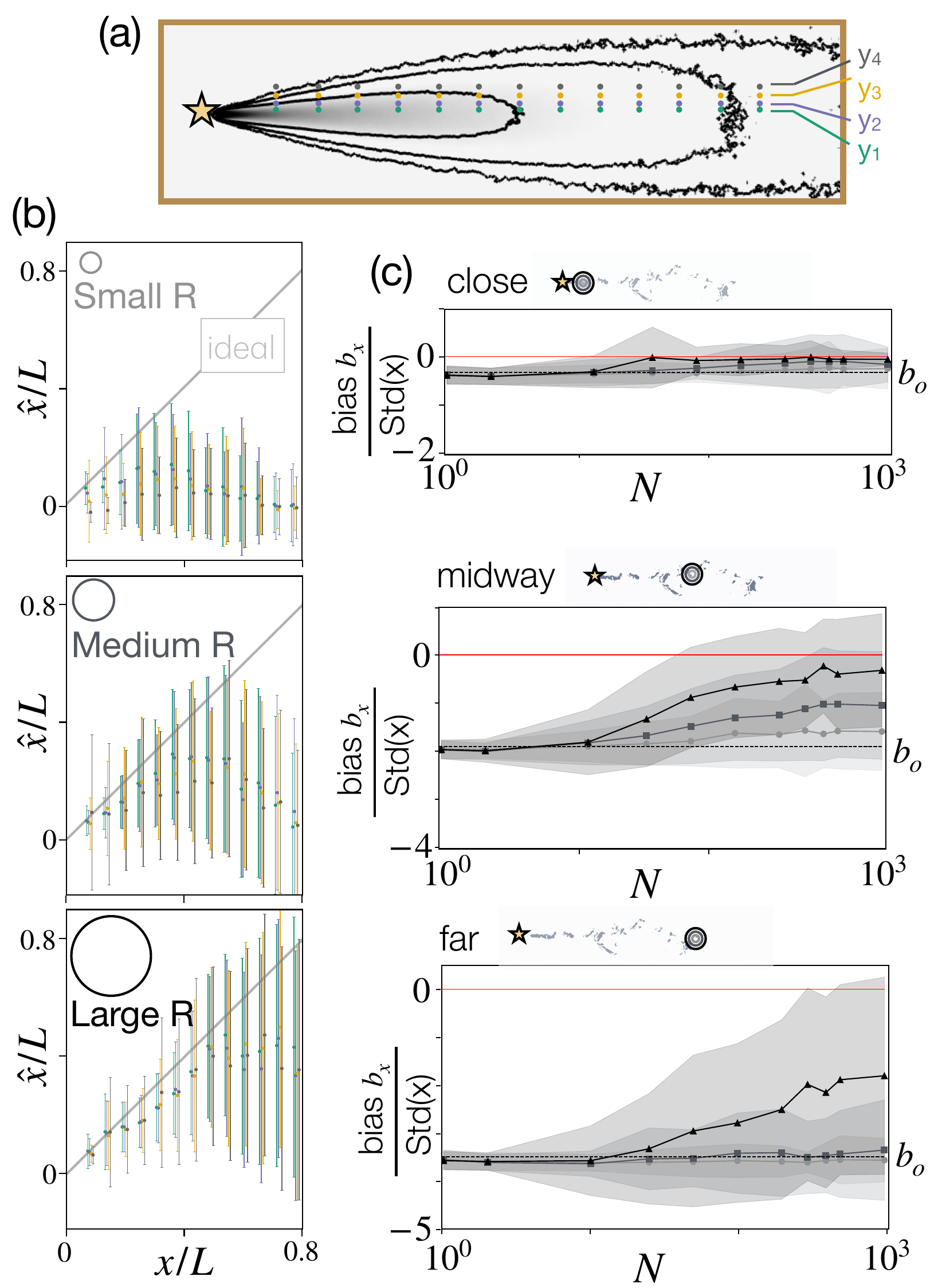} 
\caption{\label{multi_perf} Instantaneous inference with many sensors and perfect proprioception is challenging due to correlations in the odor signal. (a) Sketch of test points within the domain. (b) Maximum a posteriori estimate $\hat{x}$ \emph{vs} ground truth $x$, average and standard deviation across 200 realizations. Colors represent different locations $y$ relative to the centerline, sketched in (a). Grey line: ideal estimate $\hat{x}=x$. Accuracy degrades with $x$ as detections become rare, and improves with $R$, as the agent collects increasingly independent odor measures.
(c) Performance degrades as the agent moves away from the source. Bias $b_x=\sum_{i=1}^{N_r}(\hat{x}_i-x)/N_r$, where $N_r$ is the number of realizations, normalized with the standard deviation of $x$, Std$(x)$, as a function of the number of sensors $N$ for $x/L=0.23,0.58,0.93$ from top to bottom. Mean (solid lines) and standard deviations (shades) computed across realizations and averaged for all $4$ values of $y$.  Light, medium, and dark gray represent results with $ R=10, 25, 40$ (in units of the grid spacing $\Delta x$), respectively. Bias for single sensor $b_0=-0.327, -1.904, -3.479$, calculated analytically (see Materials and Methods).}
\end{figure}

\noindent \textbf{Single sensor}. Fluctuations, intermittency and sparsity of the odor field make odor source localization in turbulent environments a difficult problem~\cite{piro_many_2024,baker2018algorithms, durve2020collective,rigolli_learning_2022,celani2014odor,murlis1992odor,rigolli2022alternation}. 
A single binary sensor clearly provides poor instantaneous inference. 
However, the agent can improve inference using a full time series of measurements gathered by its single sensor. Note that the odor signal is temporally correlated, hence a wait time $\Delta t \ge \tau_c$ between the measurements is needed to accumulate independent information, where $\tau_c$ is the correlation time of the signal. We confirm that given a fixed number of measurements $N_T=20$, the mean square error reaches a plateau as we increase $\Delta t$ beyond about $10$ to $15\,\tau$  (Fig.~\ref{fig:setting}c), where $\tau$ is the Kolmogorov timescale. Odor cues are effectively uncorrelated at these timescales, as confirmed by the fact that the asymptotic value of the MSE at $\Delta t\ge 10\tau$ compares well with the MSE evaluated from bootstrapped data for the same number of measurements $N_T$.
 
We thus fix $\Delta t = 15\tau$ to ensure uncorrelated measures of the odor, and increase the duration of the time series. Inference improves slowly (Fig.~\ref{fig:setting}d), but the error is still of order 1 after 100 measures. For reference, 100~$\times$ 15 $\tau$ is relatively slow, e.g., it may correspond to about 2 minutes in typical conditions in the atmosphere. As above, comparison between bootstrapping and Bernoulli sampling with the same empirical likelihood used for the original signal confirms that the measures are uncorrelated. Better results may be obtained by manipulating the time series appropriately to extract other useful information from more complex features of the odor cues~\cite{rigolli_learning_2022,victor,boie_information-theoretic_2018}. \\

\noindent\textbf{Multisensor inference with perfect proprioception.}
Next, we investigate instantaneous inference using many sensors, as described at the beginning of section~\ref{sec:results}. We focus on agents located at 155 locations $\bs\xi$ arranged in a rectangular grid of 31$\times$5 positions in $x$ and $y$ (Supplementary Fig.~S1) and conduct Bayesian inference using Perfect proprioception (Eq.~\eqref{eq:L} and the first of eqs.~\eqref{prop_cases_eq}). We find that performance improves with the number of sensors $N$ and with the size $R$ of the agent, as a large $R$ further decorrelates the individual sensors  (Fig.~\ref{multi_perf}a,b). Bias improves considerably compared to the bias, $b_0$, of the individual sensor (computed analytically, dashed lines in Fig.~\ref{multi_perf}b). Note that due to correlations, improvement eventually saturates with $N$. The results are qualitatively robust to distance from the mid-line (colors in Fig.~\ref{multi_perf}a) and estimates remain quite inaccurate when the agent is further downwind from the target, even for a large radius $R$.  \\

\begin{figure}[h!]
\includegraphics[width=1\linewidth]{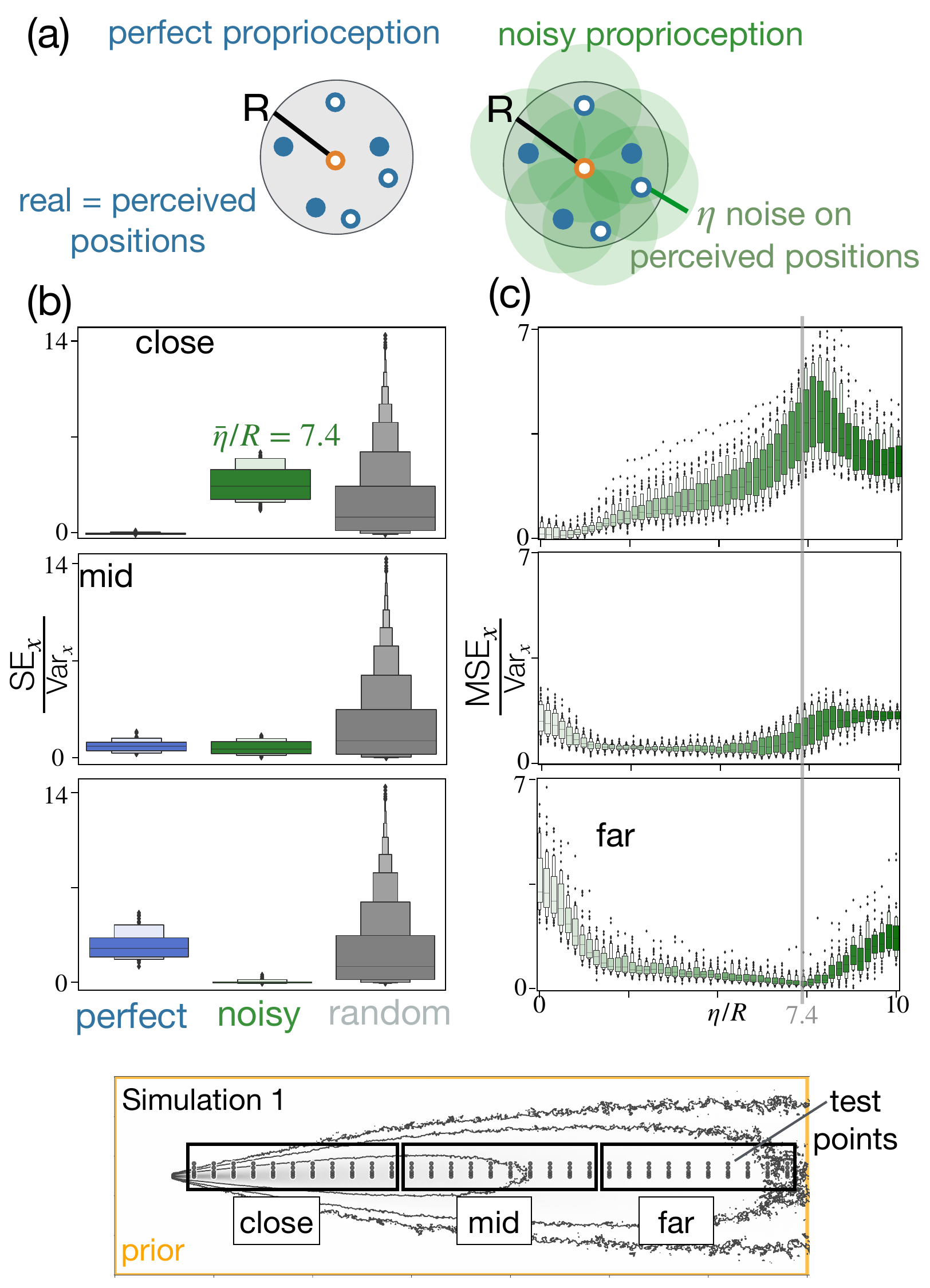} 
\caption{\label{imp_prop_fix}Proprioceptive noise improves inference far from the target. (a) Sketch of an agent with perfect proprioception (left) and noisy proprioception (right). (b) Box plot of Square error $\text{SE}=(\hat{x}_i-x_j)^2$ for $i\in(1, N_r)$ and $N_r=100$ realizations, for  55 test locations $x_j$ close to the source (top), 50 locations mid distance (center) and 50 locations far from the source (bottom). Perfect proprioception (blue), noisy proprioception with $\eta = 7.4R$ (green). Perfect proprioception outperforms noisy proprioception only close to the source and noise improves the accuracy of Bayesian estimates far from the source. A random estimate within the prior $\hat{x}_{\text{random}}\sim P(x)$ is shown for comparison (gray). (c) Box plot of mean square error $\text{MSE}_x=\frac{1}{N_r}\sum_{i=1}^{N_r}(\hat{x}_i-x_j)^2$ for the ensemble of $j$ locations close, mid, far (top to bottom), as a function of the intensity of noise ($\eta$) relative to the size of the agent ($R$) corroborating that larger values of noise are useful as the agent moves further away from the target.
}
\end{figure}

\noindent \textbf{Noisy proprioception.}
(How) Do results degrade when positional information is noisy? We focus on the same grid of real $\bs\xi$ locations and compute maximum a posteriori estimates with noisy proprioception compared to perfect proprioception (eqs.~\eqref{eq:L} and the second of~\eqref{prop_cases_eq}, sketch in Fig.~\ref{imp_prop_fix}a). We fix $N=500$, $R=25$ and $\eta=185$ in units of the grid spacing $\Delta x$. 
We first quantify error in the $x-$direction, aggregating all positions $x$ in the region closest to the target.
In this information-rich region, we find that noise is detrimental (Fig.~\ref{imp_prop_fix}b, top). But surprisingly, aggregating results for all $x$ in the central region, noise becomes irrelevant and it becomes even beneficial for positions in the furthest region (Fig.~\ref{imp_prop_fix}b, center and bottom).
To inspect whether the results depend on the strength of the noise, we repeated the simulations varying the noise from $\eta=0$ to $\eta =10 R$. We find that in the region close to the target, any level of noise is detrimental, Fig.~\ref{imp_prop_fix}c, top. In the intermediate region, noise is beneficial for most values of $\eta$, particularly for values of noise $2R<\eta<4R$ (Fig.~\ref{imp_prop_fix}c, center). The benefits of noise become pronounced within the region furthest from the target, where larger values of noise $\approx 7R$ dwarf the MSE of more than a factor 10 relative to $\eta=0$, i.e.~the case with perfect proprioception (Fig.\ref{imp_prop_fix}c, bottom).\\

\begin{figure}
\includegraphics[width=1\linewidth]{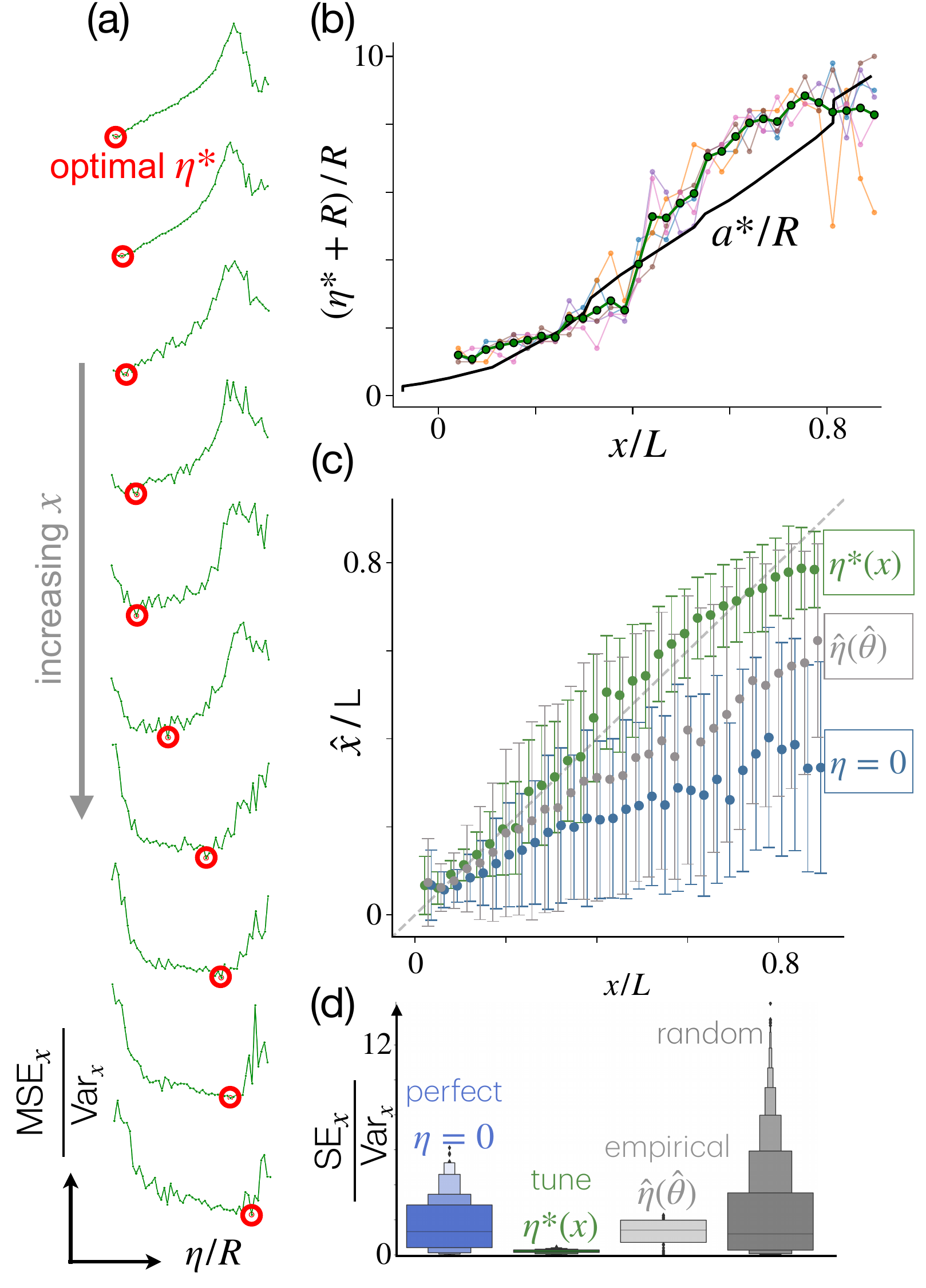} 
\caption{\label{fig:optimal_eta} 
An optimal noise exists that depends on $x$ and tuning noise to this value greatly improves inference. (a) $\text{MSE}_x$ has a marked minimum at a specific value of $\eta^*$ which increases with distance $x$. 
(b) Perceived size, at the optimal noise as a function of $x$, for different values of $y$ (pink, blue, red and orange lines), as well as their average (black). The optimal perceived size compares well with the optimal distance between detection pair, $a^*_{\text{pair}}$ (black line). 
(c) Maximum a posteriori estimate $\hat{x}$ \emph{vs} ground truth $x$, average (dots) and standard deviations (errorbars) are computed over 100 realizations and 5 values of $y$. Comparison between perfect proprioception $\eta=0$ (blue), noisy proprioception with noise tuning $\eta^*(x)$ (green) and noisy proprioception with empirical noise $\hat\eta(\hat\theta)$ (gray). 
(d) Aggregate statistics of square error across all points, color code as in panel (c). Random (darker gray): inferred position is a random point with flat probability within the prior. For these results, we excluded realizations with fewer than 2 detections, which are entirely dominated by the prior.}
\end{figure}

\noindent\textbf{Optimal noise.}
To probe this pattern systematically, we consider each point $x$ separately, rather than aggregating results for regions. For each $x$, we scan the accuracy of the inference as a function of the magnitude of positional noise $\eta$. We find that an optimal level of noise $\eta^*(x)$ exists that minimizes MSE$_x$ at each location (Fig.~\ref{fig:optimal_eta}a) and $\eta^*(x)$ systematically increases with $x$ (Fig.~\ref{fig:optimal_eta}b). If at each location the value of $\eta$ is tuned to match the local optimal noise $\eta^*(x)$, Bayesian estimates become strikingly accurate all the way through the end of the domain (Fig.~\ref{fig:optimal_eta}c).
Note that due to noise, the agent perceives its size to be larger than it is. Indeed, real sensor positions are extracted within a circle of size $R$ and noise is added on top of the real positions, hence the perceived size of the multisensor agent is $\sigma(\eta) \approx R + \eta$ and at the optimal noise:
\begin{equation}\label{eq:perc_size}
    \sigma(\eta^*) \approx R + \eta^*
\end{equation}
\noindent 
Thus, according to Eq.~\eqref{eq:perc_size}, noise increases the perceived size of the agent, which will be key to understanding why it is beneficial.  \\

\begin{figure*}
\includegraphics[width=1\linewidth]{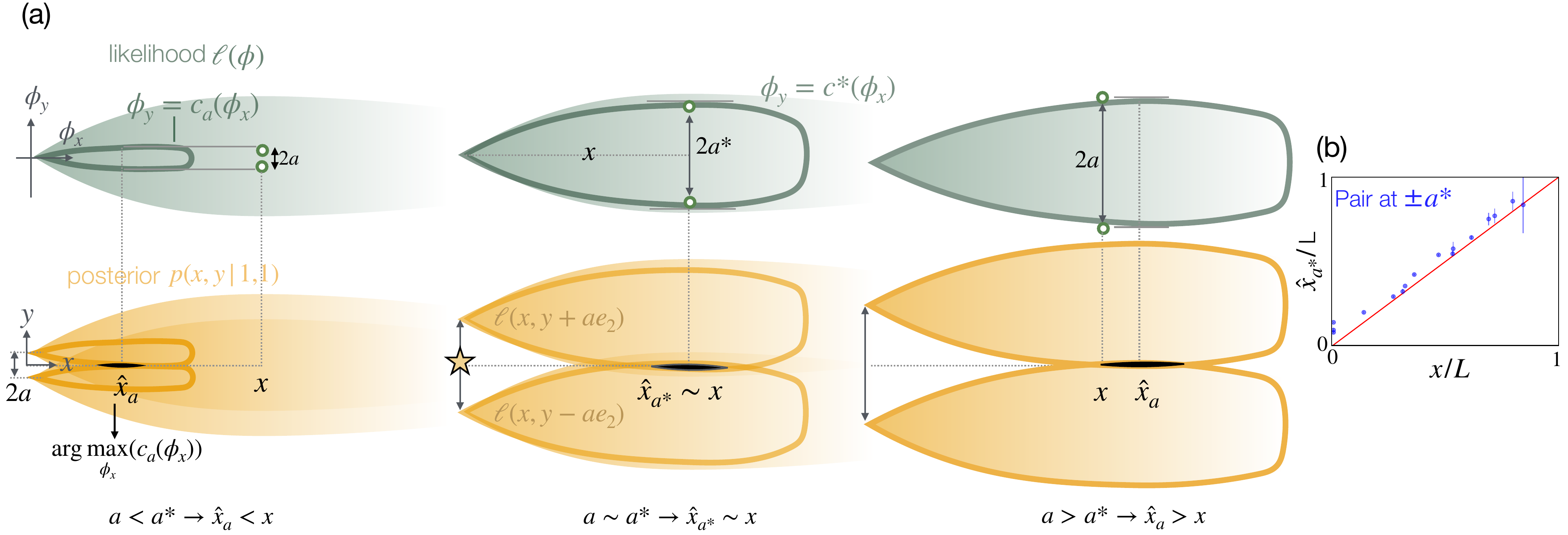} 
\caption{\label{fig:geometry} 
  The two detection limit illustrates how the perceived sensor position tunes Bayesian estimates of source location. (a) Top: Likelihood in spatial coordinates $(\phi_x,\phi_y)$ (green maps), with two detections perceived to be located at $\hat{\bm\xi}_{\pm}=(x,y\pm a)$ with $a$ small (left), intermediate (center) and large (right). The contour line of the likelihood $\phi_y=c_a(\phi_x)$ is defined as the one whose width matches the perceived pair distance: $\max c_a = a$. Bottom: posterior distribution $p(x,y|1,1)$ after measuring two detections at $\hat{\bm\xi}_{\pm}$ obtained as the product of the likelihood shifted of $\pm a$. $p(x,y|1,1)$ is maximum at the location where the contour $c_a$ is maximally wide $\hat{x}_a=\arg\max c_a$. The distance between the two detections dictates the estimated position $\hat{x}_a$ and an optimal perceived distance $a^*$ exists such that $\hat{x}_{a^*}=x$. (b) 1D maximum a posteriori estimates on the centerline of the anisotropic turbulent plume from numerical simulations with two detections separated by $a^*$ are indeed nearly perfect.
 }
\end{figure*}

\noindent\textbf{Geometric argument: the detection pair.} To rationalize the emergence of this counter-intuitive effect, we now sketch a simple geometric argument that relies on a single pair of detections (Fig.~\ref{fig:geometry}a). 
Let us consider two sensors perceived to be aligned vertically: $\hat{\bm\xi}_{\pm}=(x,y\pm a)$ 
both detecting odor, $m_+=m_-=1$. 
From Eq.~\eqref{eq:L} and the second of eqs.~\eqref{prop_cases_eq} we obtain
$(\hat{x},\hat{y})=\arg\max_{(x,y)}\ell(x,y+a)\ell(x,y-a)p(x,y)$. As long as the prior is flat, the posterior results from the product of the likelihood shifted vertically of $\pm a$ and it is maximum at the intersection of the two contour levels sketched in Fig.~\ref{fig:geometry}a, obtained directly from the contour $\phi_y=c_a(\phi_x)$ of the likelihood with its maximum at $a$: $a=\max c_a(\phi_x)$. As a consequence, the estimated position from the pair of detections is:  
\begin{equation}
\label{eq:pair}
\hat{x}_a=\arg\max_{\phi_x} c_a(\phi_x)
\end{equation}
\noindent where  $\hat{y}=0$ by symmetry. We add the subscript $a$ in $\hat{x}_a$ to emphasize that, importantly, the estimate $\hat{x}$ from Eq.~\eqref{eq:pair} depends solely on the perceived distance between the detection pair, $a$. In other words, $a$ is a knob that tunes $\hat{x}$. In particular, the further the detections are perceived to be, the larger the estimated distance $\hat{x}_a$. Clearly, an intermediate distance $a^*$ exists which makes accurate estimates $x=\hat{x}_{a^*}$. This optimum value is defined by:
\begin{equation}
\label{eq:astar}
\hat{x}_{a^*}=\arg\max_{\phi_x}(c^*(\phi_x))=x \quad {\text{;}} \quad a^*=\max(c^*(\phi_x)) 
\end{equation}
\noindent Note that the optimum distance of the detection pair, $a^*$, results from knowledge of the likelihood and of $x$, yielding the contour $c^*(\phi_x)$ as defined by eq~\eqref{eq:astar}. Using the empirical likelihood from the full simulations, we compute the optimum pair distance $a^*$ as a function of $x$ according to Eq.~\eqref{eq:astar}. The maximum a posteriori estimate $\hat{x}_{a^*}$ from a pair of detections at perceived positions $(x,y\pm \,a^*(x))$ is a nearly perfect match to $x$ (Fig.~\ref{fig:geometry}b), confirming that this geometric argument correctly captures the optimal tuning of a detection pair. But why would two detections be a relevant model for multisensor inference? \\

\begin{figure*}
\includegraphics[width=1\linewidth]{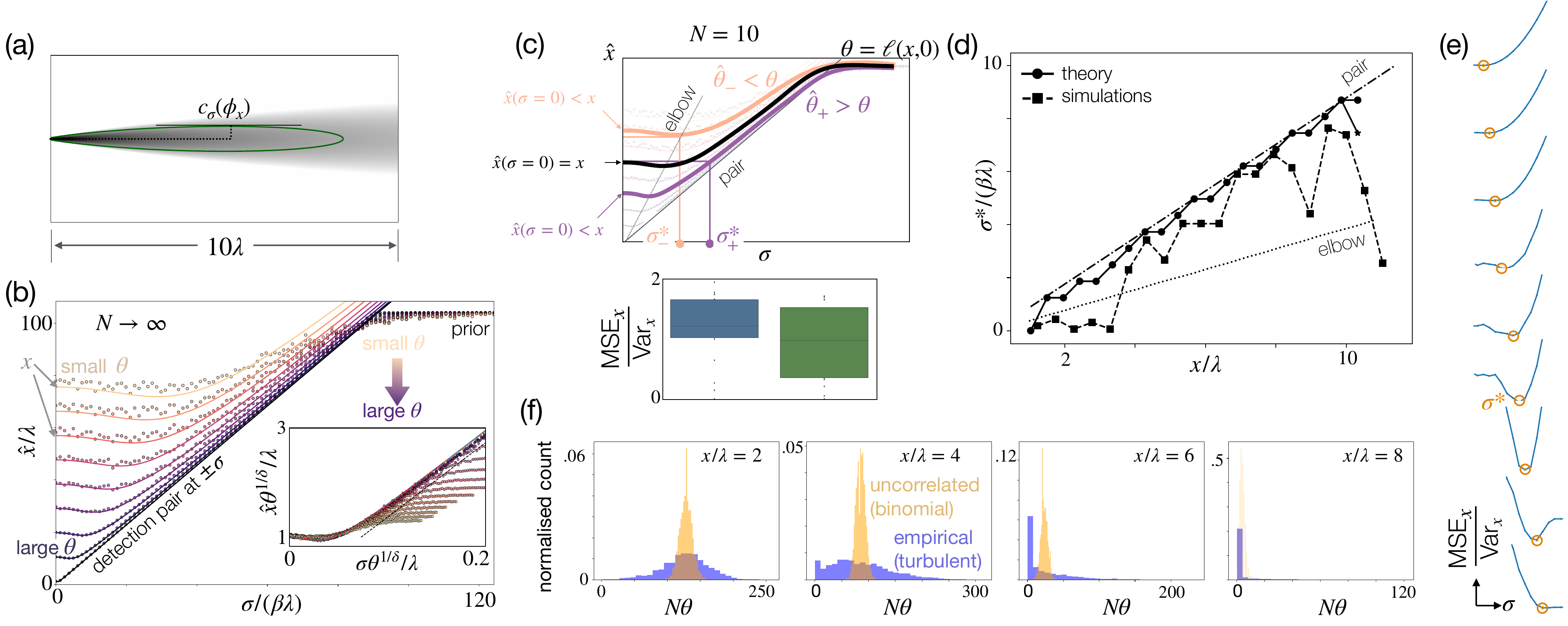} 
\caption{\label{fig:theory} Theory predicts the two-detections limit is a tight upper bound for optimal proprioceptive size. (a) Fit to empirical likelihood from fluid dynamics simulations and sketch of contour plot from two detection limits, defining the lengthscale $\lambda$ over which the likelihood decays (see text). (b) Idealized theory for maximum a posteriori estimate $\hat{x}$ from an infinite number of sensors located within a circle of size $\sigma$. Inference depends on both the perceived size $\sigma$ and the fraction of sensors that detect the odor $\theta=N_1/N$. For large $\sigma$, maximum a posteriori estimates $\hat x$ collapse onto the two-detections limit with the two sensors at distance $2a=\sigma$ (better visible within the inset). For $N\rightarrow \infty$, $\theta=\ell(x,0)$ thus inference is perfect when $\sigma=0$ (on y axis).
(c) For finite $N$, the empirical value of $\hat\theta$ differs from the likelihood, which leads to wrong estimates at $\sigma=0$ (overestimate for $\hat{\theta}<\theta$, pink, and underestimate for $\hat{\theta}>\theta$, purple). An increased perceived size $\sigma$ is beneficial to improve inference (bottom). (d) Solid lines: Optimal $\sigma^*$ for a numerical optimization of the theoretical prediction with $N=10$ sensors, where $\sigma$ is set as sketched in c. The prior only affects the furthest point (dot \emph{vs} stars). Dashed line and squares: Simulations with  $N=10$ Bernoulli sensors match the theory. (e) Maximum a posteriori from Bernoulli sampling shows that there is a well-defined optimum noise, which shifts systematically to larger values with distance (scale is omitted as irrelevant). (f) Turbulence features large departures from the expected number of detections (empirical histograms, with $N=500$). 
}
\end{figure*}

\noindent\textbf{Infinite $N$ theory for Bernoulli sampling.}
Next, we show that a group of sensors with noisy positional information behaves asymptotically as a pair of detections. 
Let us consider an infinite number of independent sensors within a small circle of radius $R$ centered at $(x,y)$ and perceived positions within a Gaussian of standard deviation $\sigma$. Values of $\sigma>R$ model the presence of positional noise, whereas $\sigma \rightarrow R$ models perfect proprioception. Let us place this group of sensors within a conical plume, with the following likelihood to detect:
\begin{equation}
    \ell(\bs\phi) = \left(\frac{\phi_x}{\lambda}\right)^{-\delta}e^{-\frac{\phi_y^2}{\beta^2 \phi_x^2}}
    \label{eq:power_law_map}
\end{equation}
\noindent represented in Fig.~\ref{fig:theory}a. Eq.~\ref {eq:power_law_map} is a canonical parametrization of turbulent odor plumes, capturing both longitudinal decay and lateral widening~\cite{celani_odor_2014}. Hence, our theoretical results apply to any anisotropic plume with similar scaling properties. The maximum likelihood estimate of source location in this idealized setting depends on two parameters: \emph{(i)} the fraction of sensors that detect odor, $\theta$ and \emph{(ii)} the perceived size of the group of sensors, $\sigma$. We consider small enough groups such that we ignore the dependence on the radius of the group of sensors $R\approx 0$. At large distances and for $N\rightarrow\infty$, the fraction of detections approximates the likelihood in the center of the group of sensors $\theta\approx \ell(x,y)$, neglecting variations of the likelihood within the group. Maximum likelihood estimates far enough from the source cannot infer $y$ and, on average, will fall on the center line. Assuming $y=\hat{y}=0$, the estimate for $\hat{x}$ asymptotically satisfies Eq.~\eqref{eq:xmin_final}, see Materials and Methods for more details on the derivation. 
Eq.~\ref{eq:xmin_final} is in quantitative agreement with numerical simulations of $N=10000$ Bernoulli sensors with $R=0$, $y=0$ and the likelihood from Eq.~\eqref{eq:power_law_map} with $\delta = 1.54$, $\lambda = 96.86$ and $\beta = 0.08$, from a fit to the empirical likelihood of our turbulent simulations (Fig.~\ref{fig:theory}a,b). The prior only affects inference in the low information limit, where most sensors detect zeros and the estimated position collapses to the furthest point within the prior. Corrections can be operated to obtain a better match to small distances -- however, this does not affect the argument.\\

For small perceived size, $\sigma\rightarrow 0$, Eq.~\eqref{eq:xmin_final}  prescribes that asymptotically the Bayesian estimates are dictated by $\theta$:
\begin{equation}
\label{eq:sigma0}
\hat{x}  = \lambda\theta^{-\frac{1}{\delta}} =x
\end{equation}
\noindent
where the second equality follows from inverting the likelihood in Eq.~\eqref{eq:power_law_map} computed in $(\phi_x=x,\phi_y=0)$. As expected $\sigma=0$ is the optimal choice and any noise is deleterious in the $N\rightarrow \infty$ limit, where information is abundant.\\ 
For large values of $\sigma$, inference degrades and the estimated location becomes asymptotically independent of sensors' readout $\theta$, but collapses onto a single curve that depends only on the perceived size $\sigma$:
\begin{equation}
\label{eq:effective_pair}
\hat{x} = \arg\max_x c_\sigma(x)=\sqrt{\frac{2}{\beta^2\delta}}\sigma
\end{equation}
\noindent
\noindent Interestingly, Eq.~\eqref{eq:effective_pair} matches Eq.~\eqref{eq:pair} for a pair of detections at perceived positions $(x,\pm\sigma)$,  assuming that the likelihood takes the form in~\eqref{eq:power_law_map}. In other words, a group of sensors on the centerline behaves effectively as a pair of detections. As a consequence, $\sigma$ plays the role of a knob that tunes the estimated position, similar to $a$ for the detection pair. 

\noindent{\bf Optimal noise tuning for Bernoulli sampling with finite $N$}. The model predicts that upon departures from these idealized conditions, noise can be tuned to improve the accuracy of the inference, as described next. For $N\rightarrow \infty$ noise provides no advantage as inference with no proprioceptive noise ($\sigma=0$) is perfect (Eq.~\eqref{eq:sigma0}, on y-axis of Fig.~\ref{fig:theory}b). Crucially, this result relies on $\theta = \ell(x,0)$, i.e.~on a fraction of detections that approximates the likelihood. However, for finite $N$ and/or correlated signals (e.g.~due to turbulence), the empirical fraction of detections $\hat\theta$ becomes a noisy version of the likelihood $\ell(x,0)$ at the real position and inference is corrupted. Estimated positions $\hat x$ from a realization where $\hat\theta>\ell(x,0)$ will underestimate $x$, as suggested by reading the idealized theory from the wrong curve (Fig.~\ref{fig:theory}c, left, purple, marked with $\hat\theta_+$). 
For these realizations, increasing the perceived size to a finite value $\sigma^*_+$ will restore the perfect inference. 
In realizations where $\hat\theta<\ell(x,0)$, inference overestimates $x$ and will weakly improve when the perceived size is near the elbow $\sigma^*_-$ (Fig.~\ref{fig:theory}c, left, pink, marked with $\hat\theta_-$). Thus, except for a narrow band of detections $\hat{\theta}\approx\ell(x,0)$, whether the group detects more or less odor than expected from the likelihood, a larger perceived size improves Bayesian estimates. The optimal perceived size is upper-bounded by the optimal distance of a pair of detections: 
\begin{equation}
\label{eq:sigmastar}
\sigma^*\lesssim a^*
\end{equation}
\noindent where $a^*$ is defined by eq~\eqref{eq:astar}. This scaling is confirmed by bootstrapping $\hat\theta$ from the binomial distribution that defines the Bernoulli process and reading the inferred position directly from the theory (Fig.~\ref{fig:theory}d, solid lines), as well as from maximum a posteriori estimates with $N=10$ Bernoulli sensors (Fig.~\ref{fig:theory}d dashed lines; Fig.~\ref{fig:theory}e). All arguments are here laid out for one-dimensional inference along the centerline. \\

\noindent{\bf Optimal noise tuning in turbulence, scaling argument}. 
Even at large $N$, the empirical fraction of detections may differ from the likelihood $\ell(x,0)$, for example, in a turbulent plume, where odor displays strong spatiotemporal correlations which affect detections by nearby sensors. 
From the asymptotics of 1D inference illustrated above, we expect that the optimal perceived size of a multisensor agent, $\sigma^*$, in turbulence scales as the optimal distance between two detections, $a^*$. 
Indeed, when $\hat\theta \gg \ell(x,0)$, many more sensors than expected detect odor and the optimal noise is expected to scale as the detection pair limit. Instances with $\hat\theta\approx 0$ contribute weakly to the optimal noise, as estimates $\hat x$ depend weakly on $\sigma$ for small $\hat\theta$ (see longer plateau for small $\hat\theta$ in Fig~\ref{fig:theory}b, eventually merging into the prior). Thus, we expect that the optimal perceived size of the multisensor agent in turbulence is dominated by large positive fluctuations, which are overrepresented in turbulence (see fat tails relative to the binomial distribution for $N=500$ uncorrelated samples of a Bernoulli variable, Fig.~\ref{fig:theory}f). 

In summary, our asymptotic theory in 1D suggests that the optimal size $\sigma^*$ of a multisensor agent in turbulence scales as the optimal distance between two detections,
i.~e.~$\sigma^* \sim a^*$. As noise modulates the perceived size $\sigma^*\sim R+\eta^*$ (see~Eq.~\eqref{eq:perc_size}), the optimal size $\sigma^*$ induces an optimal error $\eta^*$: 
\begin{equation}
    \label{eq:etastar}
    \eta^* \sim a^* - R
\end{equation}
\noindent 
where $a^*$ is defined by Eq.~\eqref{eq:astar}, from knowledge of the likelihood and $x$. We tested this scaling with our full two-dimensional inference, with noise optimized to minimize error in the estimated position $\hat x$ using the empirical likelihood from the simulations. 
Our results confirm that the optimal noise is in agreement with the expected scaling (see Fig.~\ref{fig:optimal_eta}b). 
Clearly, two-dimensional inference by a group of sensors is more complex than our geometric argument can capture. As a consequence, even at the optimal error, inference is noisier than in the one-dimensional case with the detection pair at $\pm a^*(x)$ (compare Fig.~\ref{fig:geometry}b with Fig.~\ref{fig:optimal_eta}c, green). 
However, variance increases only slightly
and the agreement between the optimal noise and our one-dimensional asymptotics Eq.~\eqref{eq:etastar} is rather satisfying (Fig.~\ref{fig:optimal_eta}b).
Interestingly, noisy proprioception allows the group to still retain its predictive power in the $y$ direction. Indeed, increasing the perceived size does not degrade estimates in the $y$ direction relative to perfect proprioception, likely because inference in $y$ is only possible in the first half of the domain closer to the source, and here the proprioceptive error remains contained $\eta^* \lesssim R$ (Supplementary Fig.~S7). In the second half of the domain, further from the source, inference in $y$ is no longer possible, and the optimal noise is larger than the size of the agent, so that the perceived positions of the sensors are nearly unrelated to their real positions. At these locations, it would be more appropriate to call the proprioceptive error a tunable size parameter. \\

\noindent
{\bf Empirical noise tuning.} So far, we discussed how noise can be tuned to optimize prediction accuracy. But optimal tuning cannot be done in practice by real biological organisms or robots, as $\eta^*$ depends on $x$, which is unknown. 
To set noise with no knowledge of $x$, we define an empirical noise 
$\hat\eta=\eta^* ( x_{\hat{\theta}} )$ where $x_{\hat{\theta}}$ is a guess for $x$ obtained from the fraction of detections $\hat\theta$, inverting the empirical likelihood along the center line (see Materials and Methods). 
The noisy proprioception inference with the empirical noise $\hat\eta$ is degraded relative to optimal tuning, yet considerably better relative to the agent with perfect proprioception, Fig.~\ref{fig:optimal_eta}c-d. Similar to what was observed for the optimally tuned proprioceptive noise, the empirical noise does not degrade the predictions in the $y$ direction relative to perfect proprioception (Supplementary Fig.~S8 and S9). 
In practice, noise pushes estimates $\hat{x}$ to larger values (Supplementary Fig.~S10), which corrects the negative bias observed for perfect proprioception.\\

\begin{figure}
\includegraphics[width=1\linewidth]{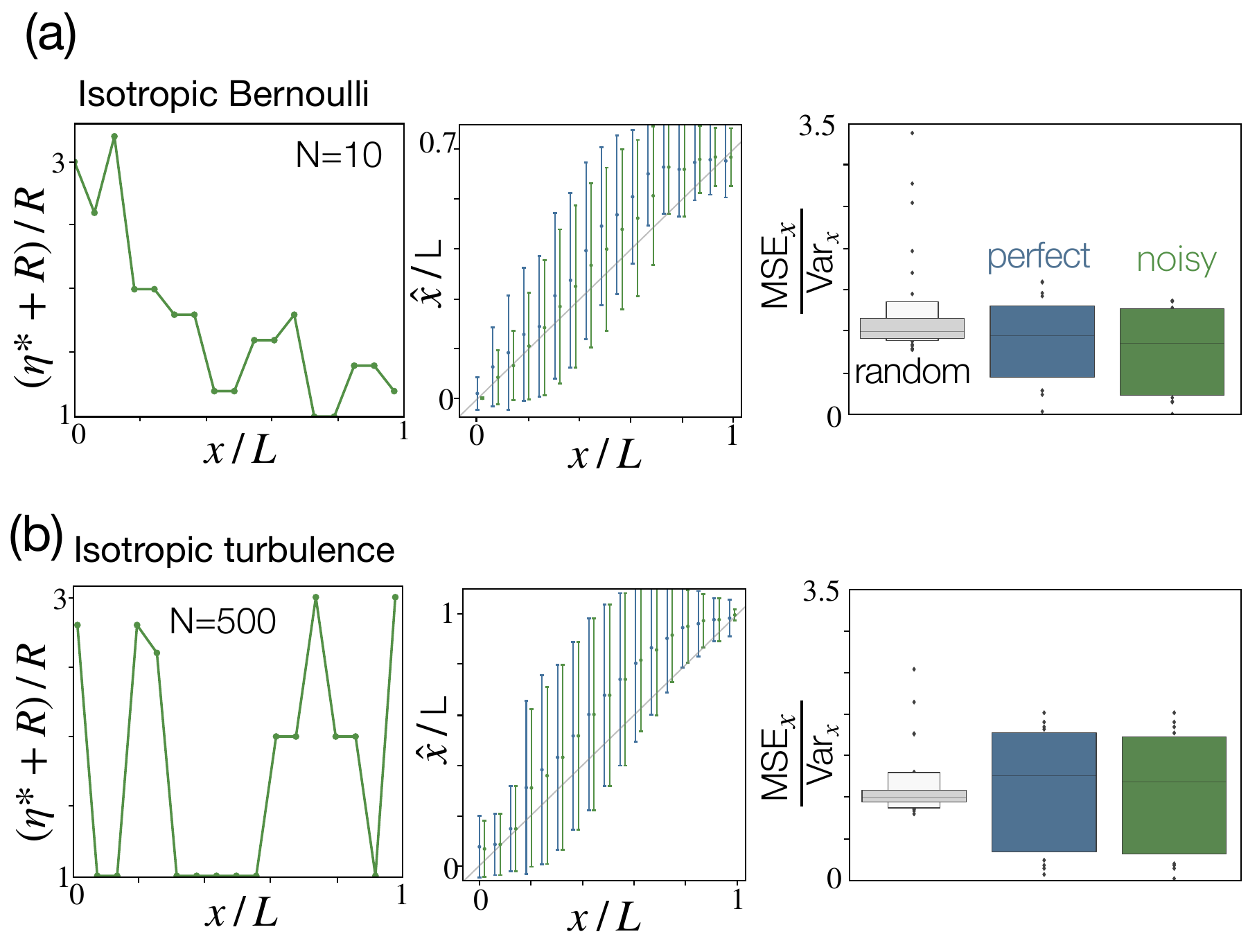} 
\caption{\label{isotropic}The benefits of noise disappear in isotropic conditions where geometry cannot be leveraged.  (a) Optimal noise (left) and estimated position \emph{vs} ground truth for $x$ (center) for isotropic Bernoulli sampling with perfect (blue) and noisy (green) proprioception. As expected, noise does not improve performance (right). To avoid washing out any benefit of noise, we use a small number of sensors $N=10$ (results for $N=500$ confirm the same trend, with a much better precision, data not shown). 
(b) Same as (a), for isotropic turbulence, confirming that the benefits of proprioceptive noise rely on anisotropy. 
}
\end{figure}

\noindent\textbf{Isotropic odor plumes.}
Our geometric argument relies on the anisotropic nature of the odor plume and is valid regardless of correlations. As we discussed above, noise in proprioception does indeed improve Bayesian inference with binary observations obtained either by thresholding anisotropic turbulent plumes (Fig.~\ref{fig:optimal_eta}c-d) or by Bernoulli sampling (Fig.~\ref{fig:theory}c, bottom). In both cases, the optimal noise follows the geometric argument sketched above (Fig.~\ref{fig:optimal_eta}a-b and Fig.~\ref{fig:theory}d-e). The effect is more prominent in the presence of correlations, which make the problem challenging; in the absence of correlations, the problem is far simpler and we use $N=10$ to avoid washing out the benefits of noise. 
On the flip side, we predict that the benefits of proprioceptive noise disappear for isotropic odor plumes, regardless of correlations. Indeed, Bayesian inference with isotropic Bernoulli and turbulent odor plumes shows no benefit of noise, compared to the case of perfect proprioception (Fig.~\ref{isotropic}). Note that, so far, noise only affects the perceived position of the sensors, but not their real positions. Hence, proprioceptive noise does not 
affect the sensory input but only its processing. This is in contrast to other sources of noise, described in the next section. \\

\begin{figure}
\includegraphics[width=1\linewidth]{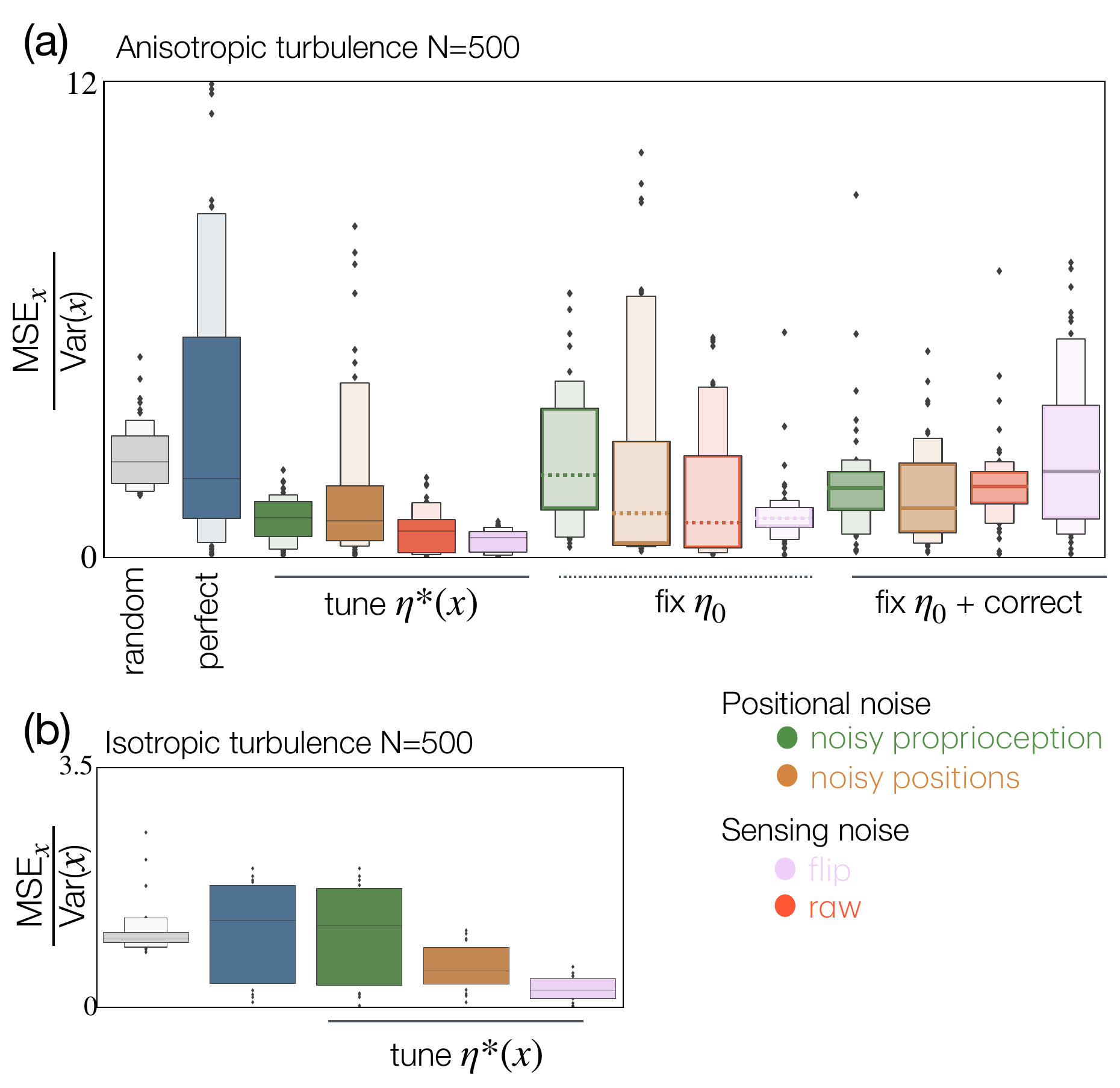} 
\caption{\label{all_err}
Distinct sources of noise all generally improve inference, even if the noise is fixed rather than tuned. (a) Systematic comparison of performance across all four noise sources and three ways to set the noise intensity, i.e.~tuning to the optimal $\eta^*(x)$ (left) using a set value $\eta_0$ (center) and fixing the noise but accounting for it within the likelihood (right).  
Box plot visualizes median (line), 25° percentile (dark box) and 75° percentiles (light box) of $\text{MSE}_x=\sum_{i=1}^{N_r}(\hat{x}_i-x_k)$, where $k$ runs over the 52 locations used for testing. 
All noise sources (colors as in legend) generally improve both relative to a random estimate flat within the prior (gray) and relative to the agent with perfect positional information and sensing (blue). Tuning the noise clearly provides the best performance, but fixed noise and corrected noise also provide good performance.
We use $\eta_0=25,25,0.001,0.01$  respectively for proprioceptive noise, positional noise, random flipping of binary signal and additive noise on the raw signal. 
Odor data and test locations from Simulation~2 (see Materials and Methods and Fig.~S5, center).
(b) Same statistics for isotropic turbulent flow (Simulation~3, see Materials and Methods and Fig.~S5, bottom).
}
\end{figure}

\noindent \textbf{Other noise implementations.}
Next, we ask whether other sources of noise may equally improve predictive accuracy. We repeat all procedures described above using a smaller dataset (see Supplementary Fig.~S1) in order to conduct a systematic test across four distinct modalities affecting either position (items 1.~and 2.~below) or sensing (3.~and 4.):
\begin{enumerate}
\item Noisy proprioception: blur the perceived position of the sensors. $\tilde{\bs{\xi}}_i = \bs{\xi}_i + \bs{\gamma}_i $ with $\gamma\sim \mathcal{N}_T(0,\eta^2)$ truncated between $(-\eta-1,\eta+1)$. \\ This is the noise we have considered so far: it affects the perception of sensor location. A byproduct of this definition is that noise increases the perceived size of the agent.
\item Noisy positions: blur the real position of the sensors, or $\bs{\xi}_i = \bs{\tilde\xi}_i + \bs{\gamma}_i $, with the same Gaussian distribution of noise as above $\gamma\sim \mathcal{N}_T(0,\eta^2)$ truncated between $(-\eta-1,\eta+1)$. With this definition, as a consequence of noise, the agent is perceived as smaller than it really is, or equivalently, the real positions lie outside the circle of radius $R$. 
\item Randomly flip binary signal: swap the binary odor observation $0\rightarrow 1$ and $1\rightarrow 0$ with probability $\eta$, or $\tilde{m}_i = m_i +\rho \,\mod2$ with $\rho\sim {\text{Bernoulli}}(\eta)$.
\item Add noise to raw signal: blur the raw odor with Gaussian noise. $\tilde{c}_i=c_i+\Gamma$, and $\Gamma\sim \mathcal{N}(0,\eta^2)$, and then threshold as described in Eq.~\eqref{eq:thresholding}.
\end{enumerate}
\noindent We repeat our Bayesian inference scheme with anisotropic turbulent plumes, optimizing noise at each location. Interestingly, the accuracy of the inference benefits from all these different ways of implementing noise, particularly from flipping the binary signal, which achieves striking accuracy (see Fig.~\ref{all_err}, Supplementary Fig.~S3). Note, however, that we optimize noise to improve the accuracy of Bayesian inference in $x$; at this optimal noise level, inferring $y$ is only possible for positional noise, whereas, for sensing noise, the estimated $\hat{y}$ invariably falls back on the centerline (see Supplementary Fig.~S4). \\

\noindent \textbf{Fixed noise and error correction.}
We ask whether noise is beneficial even if tuning is avoided altogether. 
We set the noise to a single value for all locations: $\eta_0/R=1$ for noisy proprioception and for noisy positions, $\eta_0=0.01$ for flipping noise, $\eta_0/c^0=0.001$ for noise on the raw signal. We find that all noisy agents outperform the ideal agent (Fig.~\ref{all_err}a). The constant value of $\eta_0$ is chosen to be within a range that is beneficial at least in some locations (see Supplementary Fig.~S5, note interesting non-monotonic behaviors and asymptotic plateaus for sensing noise).\\
Finally, we consider that the agent may be aware that its proprioception and sensing are noisy, and knowing the value of $\eta_0$, it may correct for it. We introduce error corrections as explained in Materials and Methods.   
We find that, for positional noise, error correction improves either bias or variance or both relative to the uncorrected estimates (Supplementary Fig.~S3). In contrast, correcting for sensing noise achieves more mixed results: for noise on the raw signal, corrections decrease bias but increase variance relative to the uncorrected case. Whereas for noise on the binary signal (flip), corrections are deleterious for both bias and variance (Supplementary Fig.~S3). \\

\noindent In the aggregate, when compared to the ideal agent that has perfect proprioception and sensing, all four noise generation mechanisms sensibly improve accuracy, whether the noise level is (unrealistically) tuned to optimize performance or it is fixed, or it is fixed and corrected for. This is seen by the aggregate statistics of MSE$_x$ over all 52 test locations (Fig.~\ref{all_err}a). The most striking aggregate results are obtained for the fixed flipping noise, which dwarfs the average $\langle\text{MSE}\rangle_x$ from $3.5$ to $0.4$. 

The reason why other sources of noise also improve inference lies at least partially in their ability to break correlations. Indeed, if noise affects the position where sensors really are (noisy positions, item 2), then with our definition of noise, they will sample outside of where they are perceived to be. Their measures are then expected to be less correlated than they would be if they were sampling at points that were closer to one another. Sensing noises (flip and raw) also break correlations as they introduce uncorrelated mistakes in the measures. We thus expect that these noise mechanisms will succeed in the presence of correlations, whether the signal is isotropic or anisotropic (consistent with inference using turbulent odor plumes~\cref{all_err}a,b) and fail in the absence of correlations (consistent with results using isotropic Bernoulli sampling Supplementary Fig.~S6; note a residual benefit of flipping for anisotropic Bernoulli sampling, potentially related to geometry or to the well studied threshold effect of stochastic resonance~\cite{carroll2006measurement}). 

\section{Discussion}

Inspired by the octopus, we analyzed the problem of inferring the source of a turbulent odor plume, based on detections from many sensors distributed in space. Most of our results are dedicated to understanding the role of proprioceptive noise: sensors are still distributed uniformly within a circle of radius $R$, but their perceived location is blurred with respect to their real location. We find that an appropriately tuned level of proprioceptive noise improves inference. This counterintuitive effect emerges because, as a result of noise, the agent perceives it to be bigger. If properly controlled, this allows to leverage additional information hidden within the geometry of the odor plume. Indeed, in the presence of a net fluid flow, the odor gets transported within a conical region of space, and Bayesian estimates depend on this anisotropic geometry. We show that asymptotically, the group of sensors behaves like an effective pair of detections; the perceived size of the agent corresponds to the perceived distance between the two sensors, which can be tuned to make more accurate estimates of source position. The theory is developed and tested for Bernoulli sampling, in which case the model perfectly captures the statistics of the problem. In this case, Bayesian inference is guaranteed to converge asymptotically~\cite{bayesian_convergence} and the benefits of noise are only visible for small $N$, far from the asymptotic limit. 

Interestingly, while the theory is developed for Bernoulli detections, the additional information provided by geometry persists even in the presence of spatiotemporal correlations caused by turbulence. In fact, because the problem becomes significantly more challenging, the benefits of noise are much more prominent in realistic turbulent plumes rather than in idealized Bernoulli sampling. Consistent with the key role of geometry put forward by the theory, we find that noise fails in isotropic odor plumes, both for Bernoulli sampling and turbulence. Note that in the presence of turbulence, the model is misspecified, as it does not account for the strong correlations between odor detections. However, proprioceptive noise leaves all correlations unaltered; it cannot cure this inaccuracy of the model. Instead, it leverages geometric information that is not otherwise exploited by the model. To our knowledge, this mechanism has not been described before.

The theory also predicts that the optimal strength of proprioceptive noise depends on the location of the agent relative to the target, in agreement with the numerical results. Clearly, real organisms and robots cannot tune noise to this optimal level, as they do not know the location of the target. However, agents may leverage a scaling argument to estimate the optimal noise based on current observations only. For this empirical noise tuning, the likelihood of odor detections is used twice: once for Bayesian inference, and once to compute the optimal two-detection distance from geometry. To forfeit knowledge about the extension of the odor plume, further work may extend our approach using model-free supervised algorithms that approximate Bayesian inference (see e.g.~\cite{rigolli_learning_2022}), or by approximating the shape of the plume from local measurements of average flow speed and fluctuations. \\
Thus, in the aggregate, our results demonstrate that noisy proprioception does not hinder accurate inference. In fact, approximate noise tuning functions well in realistic turbulent conditions, suggesting that noisy proprioception in robotics may even be tuned to improve inference based on sensory information.\\

\noindent Octopus inspired robotics is receiving considerable attention~\cite{giordano2021perspective,xie2023octopus}. Precise proprioception is often deemed necessary for complex motor control in soft robotics~\cite{wang2018toward}, and the challenges of encoding for the shape of a soft body plan have been recently addressed with deep learning algorithms~\cite{truby2020distributed}. Further work is needed to test whether specific tasks may be best accomplished with noisy proprioception, thus avoiding the need for acquisition and processing of accurate proprioceptive feedback from integrated soft sensors, which require advanced hardware and software~\cite{kim2021review,wang2018toward}. 

\noindent In living organisms, olfactory processing may be affected by proprioceptive noise. 
The brain has long been hypothesized to perform Bayesian inference and theory and experiments suggest that noise may offer advantages to both detect weak sensory input as well as encode the signal probabilistically, which can facilitate Bayesian inference (see e.g.~\cite{noda2019bayesian, knill2004bayesian},  
\cite{ma2006bayesian,knill2004bayesian,li2017neural,ermentrout2008reliability} and references therein). Computational and experimental work suggests noise may be beneficial to process sensory information, see e.g.~\cite{moss2004stochastic,mcdonnell2009stochastic}. \\
\noindent For octopus, knowledge is currently insufficient to test whether noise affects the processing of distributed chemosensory information. 
Behavioral experiments show that the brain does process sensory information collected by the arms~\cite{gutnick2020use,zullo2011new, hochner2013nervous}. However, to what extent proprioceptive information is available and is integrated into sensory processing and motor control remains to be established~\cite{zullo2025}. 

Unfortunately, the molecular mechanisms for proprioception in octopus remain elusive~\cite{TUTHILL2018R194, wells2013octopus, gutnick2011octopus,zullo2025}, thus
targeted perturbation of proprioception, as available now for e.g.~insects~\cite{mamiya2018neural, oliver2021molecular},
is unfeasible for an octopus. 
To quantify proprioception, indirect evidence may be collected by testing behavior, e.g., letting the octopus solve an arm maze with no vision, or searching, moving objects, and 
manipulating prey with no visual cues~\cite{gutnick2020use, gutnick2011octopus, flash2023biomechanics, nesher2014self, crook2014neuroethology, sumbre2005motor, sivitilli2023mechanisms}. 
Moreover, further work is needed to understand how chemosensory information is interpreted, particularly whether each sucker processes its local sensory information or rather all information is pooled and then processed centrally. \\

Results with other noise sources can be rationalized as means to correct a misspecified model. Indeed, when noise affects the sensing process itself, or the real position of the sensors, it effectively breaks correlations in the odor, partially correcting the model assumption of independent observations. Consistently, the positional noise on real sensor location is only beneficial in the presence of spatiotemporal correlations caused by turbulence, and not for Bernoulli sampling, regardless of geometry. Similarly, the flipping noise is beneficial in turbulence and not for Bernoulli isotropic sampling, although it has beneficial effects in anisotropic Bernoulli as well, suggesting it may leverage both correlations and geometry. Note that all results in the presence of turbulence are obtained for a large number of sensors $N=500$, consistent with the fact that the problem is much more challenging, i.e., asymptotic convergence for Bayesian inference is not guaranteed when the model is misspecified~\cite{bayesian_misspecified}. Other ways to correct for the misspecification of the model were discussed for turbulent source localization in~\cite{piro_many_2024}.\\
Well beyond the specific problem of turbulent source localization, the idea that noise may improve inference has a long history in information processing (see e.g.~\cite{adaptive_stoch_resonance,stoch_resonance_revmodphys}) and is often referred to as stochastic resonance after ref.~\cite{benzi_sutera_vulpiani}). Noise-enhanced Bayesian estimators can be proven theoretically under assumptions on the statistics of the signal (see e.g.~\cite{CRLB_noise}). Further work is needed to adapt these arguments to the complex statistical properties of turbulent odor plumes~\cite{celani_odor_2014, duplat2008mixing, shraiman2000scalar, villermaux2003mixing, orsi2021scalar}. Another interesting avenue for further theoretical work is the connection between Bayesian inference with empirical loss minimization, where, under assumptions on the model and the structure of the data, noise on labels has been shown to play the role of a regularizer~\cite{bishop}.\\

The counterintuitive benefits of noise for processing turbulent information represent an exciting avenue for further research and may lead to novel concepts of adaptation to uncertain environments in living systems as well as optimization principles in robotics. 

\begin{acknowledgments}
This research was supported by grants to AS from the European Research Council (ERC) under the European Union’s Horizon 2020 research and innovation programme (grant agreement No 101002724 RIDING) and the National Institutes of Health (NIH) under award number R01DC018789. The European Commission and the other organizations are not responsible for any use that may be made of the information it contains. We thank Francesco Viola for sharing a GPU-accelerated version of the CFD code, as well as support and discussions regarding computational fluid dynamics.
\end{acknowledgments}

\begin{appendices}

\section{Inference scheme} \label{Appendix A}

Multisensor Bayes' inference relies on computing the posterior distribution of parameters given observations:
\begin{equation*}
    P(\bs{\xi}|\bs{m}) \propto L(\bs{m}|\bs{\xi}) \cdot p(\bs{\xi})
\end{equation*}
\noindent where $P(\bs{\xi}|\bs{m})$ is the posterior distribution, $\bs{m}$ is the array of binary measurements made by the $N$ sensors and $\bs{\xi}=(x,y)$ is the set of parameters to be inferred, i.e.~the location $(x,y)$ of the center of the agent, relative to the target. $L(\bs{m}|\bs{\xi})$ is the likelihood and $p(\bs{\xi})$ is the prior, which we consider uniform and flat in a rectangular region depicted in Fig.~\ref{fig:setting}b. We discard the evidence term in the denominator in Bayes' theorem, since we are not interested in model comparison. We assume independent measures and thus factorize the likelihood $L(\bs{m}|\bs{\xi}) =  \prod_{i=1}^N L_{i}(m_i|\bs{\xi})$, where $L_{i}(m_i|\bs{\xi})$ is the likelihood that the $i-$th sensor measures $m_i$ given the center is at $\bs\xi$. Clearly, the model is misspecified as measures are not independent due to spatiotemporal correlations inherited by turbulence~\cite{heinonen2025exploring}. However, turbulent correlations are notoriously hard to model and we will therefore stick with this simplifying assumption (see e.g.~\cite{piro_many_2024, celani_odor_2014}). Note that noise enhances Bayesian estimates for both correlated and uncorrelated signals, and our asymptotic analysis shows why the effect is even more pronounced for correlated signals, as we rationalize in the main text. \\
We define 
\begin{equation}\label{lkl_eq}
    L_{i}(m_{i}|\bs{\xi}) =\int \ell_{i}^{m_{i}} (1-\ell_{i})^{1-m_{i}}p(\bs{\phi}_i|\bs{\xi}) d\bs{\phi}_i
\end{equation}
where $\ell_i=p(m=1\:|\:\bs{\phi_i})=\ell(\bs\phi_i)$ is the probability that a sensor measures $m=1$ given that it is located as $\bs\phi_i$ and is purely dictated by physics. $p(\bs{\phi}_i|\bs{\xi})$ is the probability to find sensor $i$ at location $\bs\phi_i$ given that the center is at $\bm\xi$, and models proprioception. We identify three models of proprioception:

\begin{equation}\label{prop_cases_eq_matmet}
p(\bs{\phi}_i|\bs{\xi}) = 
    \begin{cases} 
            \delta (\bs{\phi}_i - \bs{x}_i(\bs{\xi})) \: \: \: \text{Perfect}\\
       \delta (\bs{\phi}_i - \tilde{\bs{x}}_i(\bs{\xi})) \: \: \: \text{Imperfect Unaware}   \\
       \mathcal{N}_T(\tilde{\bs{x}}_i(\vec{\xi}), \eta^2)\: \: \: \text{Imperfect Aware}         
        \end{cases}
\end{equation}
\noindent Perfect proprioception is characterized by a Dirac delta centered on the real positions of the sensors relative to the center: $\bs x_i(\bs \xi)$. The next two lines model two different cases of imperfect proprioception. The Imperfect Unaware agent ignores positional noise: the Dirac delta function is centered at the wrong positions $\tilde{\bs x}_i(\bs \xi) \sim \mathcal{N}_T(\bs x_i(\bs \xi), \eta^2)$, where $\mathcal{N}_T$ denotes the truncated normal distribution that generates positional noise. 
In the Aware scenario, the agent uses the correct proprioceptive distribution, and the integral in Eq.~\ref{lkl_eq} effectively corrects for positional noise. \\
To obtain a single point estimator, we select the \textit{maximum a posteriori} (MAP)~\cite{Marcelo, Clason, Wacker2020MAPEF, Helin_2015, dashti2017bayesian, Burger_2014, bassett2019maximum,GRIBONVAL202149,tarantola2005inverse}:
\begin{eqnarray}
    \hat{\bs \xi}  
    = \arg \max_{\bs \xi} L(\bs{m}|\bs{\xi}) \cdot p(\bs{\xi})
\end{eqnarray}

We numerically compute the MAP through the TPE (Tree-Structured Parzen Estimator) algorithm using the Python library Optuna~\cite{DBLP:journals/corr/abs-1907-10902}. To obtain the statistical quantities, we repeat the entire process $M$ times. Hence, we can compute the ensemble mean of the estimated position $\hat{\bs\xi}$ and its standard deviation:

\begin{equation}\label{estimatiom_r}
\langle\hat{\bs \xi} \rangle=\frac{\sum_{j=1}^M \hat{\bs \xi}_j}{M}
\end{equation}

and:

\begin{equation}\label{var_r}
StD(\hat{\bs \xi})=\sqrt{\frac{\sum_{j=1}^M (\hat{\bs \xi}_j-\langle \hat{\bs \xi}\rangle)^2}{M}}
\end{equation}

\noindent The standard deviation quantifies the dispersion of the estimates $\hat x$ across the $M$ realizations, indicating the robustness of the inference process. All figures visualizing estimated target location \emph{vs} ground truth represent $ \langle\hat{\bs \xi}\rangle \pm StD(\hat{\bs \xi})$

\subsection*{Performance metric}
We summarize here the notation used throughout the paper to represent different variants of the classic performance metric based on the square error between the estimate and the ground truth:
\begin{itemize}
\item Square error: $\text{SE}_x=(\hat{x}_i-x_k)^2$, where index $i$ runs over $N_r=100$ for the large dataset, $N_r=200$ for the small dataset realizations and $k$ runs through the 155 test locations for the large dataset and 52 for the small dataset represented in Supplementary Fig.~S1

\item Mean square error: $\text{MSE}_x=1/N_r\sum_{i=1}^{N_r}(\hat{x}_i-x_k)^2$, where $k$ runs through the  155 test locations for the large dataset and 52 for the small dataset represented in Supplementary Fig.~S1

\item Average mean square error: $\langle \text{MSE}_x \rangle=1/(N_rn_y)\sum_{k=1}^{n_y}\sum_{i=1}^{N_r}(\hat{x}_i-x_k)^2$, where $k$ now runs over the $n_y$ test points with a fixed $x$ and different $y$. $n_y = 5$ for the large dataset, $n_y=4$ for the small datset.

\item Positional variance: $\text{Var}_x=1/N_p\sum_{k=1}^{N_p}(\bar x - x_k)^2$, where $N_p$ is the number of test locations and $\bar x = 1/N_p \sum_{k=1}^{N_p}x_k$. This variance is used as a normalization factor. 

\item All the previous definitions hold for the $y$ coordinate as well, with the substitution $x \leftrightarrow y$.

\item The bias $b_0$ is analytically calculated and it depends on the prior. 
\begin{center}
$b_0 = (x_0 \cdot p_0(\bs{\xi}) + x_s \cdot p_1(\bs{\xi}))-\bs{\xi}$,
\end{center}
where $x_0$ is the average position of convergence of the MAP when the sensor detects $0$. $x_0$ strongly depends on the prior choice. $p_0(\bs{\xi})$ is the probability to detect $0$ at a given position $\bs{\xi}$, $x_s$ is the source position and $p_1(\bs{\xi})$ is the probability to detect $1$ at position $\bs{\xi}$. The probability is computed as an average over the different $y$s used.
\end{itemize}

\begin{table*} [t]
\setlength\tabcolsep{4pt}
\centering
\begin{tabular}{|c | c| c| c| c| c| c| c| } 
\hline
Figures & $N$ & $\lambda$ & $L\times H$ & $N_p$ & $N_r$ & $R$ & \\
\hline
1  & 500 & 96.86 & 863 $\times$ 94 & 52 & 200 & - &Simulation 2 \\
2 & 500 & 96.86 & 863 $\times$ 94 & 52 & 200 & 10, 25, 40&Simulation 2 \\
3 and 4 & 500 & 176.76 & 1750 $\times$ 500 & 155 & 100 & 25& Simulation 1 \\
5 & 2 & 96.86 & 863 $\times$ 94 & 52 & 200 & - & Simulation 2 \\
6 & 10 & 96.86 & 990 $\times$ 1 & 20 & 100 &  - & Bernoulli, likelihood  Eq.~\eqref{eq:power_law_map} with $\lambda$ from fit to Simulation~2\\
7a & 10 & 35.27 & 165 $\times$ 1 & 17 & 100 & 25& Bernoulli, empirical likelihood Simulation~3  \\
7b & 500 & 35.27 & 165 $\times$ 1 & 17 & 100 & 25& Simulation 3  \\
8 & 500 & 96.86 & 863 $\times$ 94 & 52 & 200 & 25& Simulation 2  \\
\hline
\end{tabular}
\caption{Parameters of the numerical inference. Reference figure, Number of sensors, representative lengthscale of likelihood $\lambda$ from fit of empirical likelihood to Eq.~\eqref{eq:power_law_map}, extent of the prior $L\times H$, number of test positions $N_p$, number of realizations $N_r$, odor observations. The likelihood is the empirical average of all snapshots in the simulation, except for Fig.~\ref{fig:theory} where the likelihood follows Eq.~\eqref{eq:power_law_map} with the corresponding parameter $\lambda$. }
\label{tab:inference_parameters}
\end{table*}

\subsection*{Error correction}
For proprioceptive noise, corrections are achieved by integrating the likelihood over all values of perceived locations (third of eqs.~\ref{prop_cases_eq_matmet}). Similarly, we introduce error corrections for the other noise modalities as an integral over the noise distribution, in line with the classical error correction approach~\cite{carroll2006measurement}:
\begin{equation}
    \bar{f}(x) = \int f(\tilde{x} - n) \, p(n) \, dn,
    \label{eq:error_correction}
\end{equation}
where $f(x)$ is a generic function of the signal $x$, $\tilde{x}$ is the noised signal, $\bar{f}(x)$ is the corrected function, $p(n)$ is the probability density function (pdf) of the noise, and $n$ the noise random variable.

For the case where the noise is added onto the raw signal, we write

\begin{equation}
L_i(m_i, \boldsymbol{\xi})
= \int
p\!\left(
    m_i  \,\big|\,
    \boldsymbol{\xi},\:
    \Theta(\tilde c_i - n - t_r)
\right)
\, p(n)\, dn ,
\label{eq:aware_likelihood_signal}
\end{equation}

where $\tilde{c}_i$ is the noisy signal, $tr$ the threshold, $\Theta(\cdot)$ 
the Heaviside step function, and the integration effectively evaluates the 
likelihood over the corrected signal $\tilde{c}_i - n$. \\
In the scenario where the noise acts as a Binary Symmetric Channel, i.e., it can flip the result of the measurement, we have:

\begin{equation}
\begin{aligned}
L_i(\tilde m_i, \boldsymbol{\xi})
&= \sum_{{m}_i \in \{0,1\}}
p\!\left(
    \tilde m_i \,\big|\,
    {m}_i
\right)
\, p\!\left(
    {m}_i \,\big|\,
    \boldsymbol{\xi}
\right)\\
&= (1-\eta)\,p(m_i=\tilde m_i \mid \boldsymbol{\xi})
    + \eta\,p(m_i=1-\tilde m_i \mid \boldsymbol{\xi}),
\end{aligned}
\label{eq:aware_likelihood_flipping}
\end{equation}

where $\tilde{m}_i$ is the (possibly) flipped signal, and the flipping process 
follows $p(\tilde{m}_i \neq m_i) = \eta$.

\section{Theory}
Let $P$ sensors, located at $\{(x_p,y_p)\}$ relative to the center $(x,y)$, detect a nonzero signal and $Q$ sensors located at $\{(x_q,y_q)\}$ detect zeroes. Under these assumptions, the total likelihood, starting from Eq.~\eqref{eq:power_law_map}, is
\begin{eqnarray}
    L(\hat x,\hat y) &=& \prod_{p=1}^P \left( \frac{\hat x + x_p }{\lambda}\right)^{-\delta}
    \exp\!\left[-\frac{\left( \hat y + y_p \right)^2}{ \beta^2\left(\hat x + x_p\right)^2}\right] \\
    &\times& \prod_{q=1}^Q \left( 1 - \left( \frac{\hat x + x_q }{\lambda}\right)^{-\delta}
    \exp\!\left[-\frac{\left( \hat y + y_q \right)^2}{ \beta^2\left(\hat x + x_q\right)^2}\right]\right) , \nonumber
    \label{eq:likelihood}
\end{eqnarray}
where $(\hat x,\hat y)$ is the estimated position of the center $(x,y)$. Since we are interested in estimates far away from the source, we assume $\hat x \gg |x_{p,q}|$ so that $\hat x + x_{p,q} \approx \hat x$. In addition, we fix the center along the centerline, i.e., $y = \hat y = 0$.

Using these approximations, the likelihood simplifies to
\begin{eqnarray}
    &&L(\hat x) =\nonumber \\
    &&\left( \frac{\hat x }{\lambda}\right)^{-\delta P}
    \exp\!\left[-\sum_{p=1}^P\frac{y_p^2}{ \beta^2\hat x^{2}}\right]
    \prod_{q=1}^Q \left(1- \left( \frac{\hat x}{\lambda}\right)^{-\delta}
    \exp\!\left[-\frac{y_q^2}{ \beta^2\hat x^{2}}\right]\right). \nonumber
\end{eqnarray}

Maximizing the above likelihood implies minimizing the exponent $S(\hat x)$ (after inverting the sign):
\begin{eqnarray}
    \label{eq:likelihood_sum}
    S(\hat x) &=& \delta P \log\!\left(\frac{\hat x}{\lambda}\right)+\sum_{p=1}^P\frac{y_p^2}{ \beta^2\hat x^{2}} \\ 
    &-&\sum_{q=1}^Q \log\!\left(1- \left( \frac{\hat x}{\lambda}\right)^{-\delta}
    \exp\!\left[-\frac{y_q^2}{ \beta^2\hat x^{2}}\right]\right).\nonumber
\end{eqnarray}

Further assuming that $y_p$ and $y_q$ are drawn from independent identical normal distributions with standard deviation $\sigma$, Eq.~\eqref{eq:likelihood_sum} can be simplified (after dividing by the total number of sensors), resulting in
\begin{eqnarray}
     \label{eq:likelihood_sum_simplified}
     s(\hat x) &=& \theta\left(\delta \log\!\left(\frac{\hat x}{\lambda}\right) +\frac{\sigma^2}{\beta^2\hat x^{2}}\right) \\ 
     &-& (1-\theta)\log\!\left(1- {\left(\frac{\hat x}{\lambda}\right)}^{-\delta}
     \sqrt{\frac{\beta^2\hat x^{2}}{\beta^2\hat x^{2} + 2\sigma^2}}\right),\nonumber
\end{eqnarray}
where $\theta = \frac{P}{P+Q}$. Minimizing the above expression for $\hat x$ gives:
\begin{eqnarray}
\label{eq:xmin_final}
        & &\theta \left( \delta \hat x^{-1} - 2\sigma^2 \beta^{-2}\hat x^{-3} \right) =\\
        & &\frac{1 - \theta}{1- {\left(\frac{\hat x}{\lambda}\right)}^{-\delta}
        \sqrt{\frac{\beta^2\hat x^{2}}{\beta^2\hat x^{2} + 2\sigma^2}}}
        \left(\frac{\hat x}{\lambda}\right)^{-\delta}
        \frac{\delta\beta^2\hat x^{2}+2\sigma^2\left(\delta - 1\right)}{\left(\beta^2\hat x^{2}+2\sigma^2\right)^{\frac{3}{2}}}\,\beta. \nonumber
\end{eqnarray}

\subsection*{Empirical noise tuning}
In the main text, we have discussed the results of inference using a proprioceptive noise $\hat\eta$, tuned using the empirical fraction of detections, $\hat \theta$. 
To define $\hat\eta$, we start from the empirical likelihood $f(x,y)=T(c(x,y)>c_{thr})/T$ corresponding to the portion of time the odor is above threshold in the direct numerical simulations of the turbulent plume. We fit these data to the likelihood in Eq.~\ref{eq:power_law_map}, and obtain the values for the parameters $\lambda,\beta,\delta$, mentioned in the main text. 
On the centerline:
\begin{equation}
\label{eq:fit_l}
    \theta = \left( \frac{x}{\lambda} \right) ^{-\delta}
\end{equation}
Plugging this expression into Eq.~\ref{eq:effective_pair}, we obtain:
\begin{equation}
    \sigma = \sqrt{\frac{\delta}{2}} \frac{\beta \lambda}{\theta^{\frac{1}{\delta}}}
\end{equation}
Using this expression with $\hat \theta$ instead of $\theta$ gives us an estimate of the optimal perceived size given the empirical measurements.
The empirical noise is thus set by:
\begin{equation}
    \hat\eta = \frac{1}{2} \sqrt{\frac{\delta}{2}} \frac{\beta \lambda}{\hat\theta^{\frac{1}{\delta}}} - R
\end{equation}
\noindent The empirical noise $\hat{\eta}$ corresponds to the optimal noise from Eq.~\eqref{eq:etastar}, with $x\equiv\hat{x}_{\hat\theta}=\lambda \hat\theta^{-1/\delta}$, which results from inverting Eq.~\eqref{eq:fit_l}.\\

\section{Fluid Dynamics Simulations}

\subsubsection*{Anisotropic turbulence}
The fluid field is modeled using direct numerical simulations of the Navier-Stokes equations in an open channel, using a semi-implicit, finite-difference scheme:
\begin{align}
\label{eq:ns}
    \rho\left(\frac{\partial \bs{u}}{\partial t} + \bs{u} \cdot \nabla \bs{u}\right) &= -\nabla P + \mu \nabla^2 \bs{u}; \\
    \nabla\cdot\bs{u}&= 0. \nonumber
\end{align}
Here, $\bs{u}$ is the fluid velocity field, $\rho$ the density, $P$ the pressure and $\mu$ the viscosity. To model the odor transport, we use an advection-diffusion equation:
\begin{equation}
\label{eq:odor}
    \frac{\partial c}{\partial t} + \bs{u} \cdot \nabla c= D \nabla^2 c,
\end{equation}
where $D$ is the diffusivity of the odor. More details about the turbulent channel flow simulations can be found in ref.~\cite{10.7554/eLife.102906} (Table~3, Simulation ID 2 and 3). Simulation parameters are presented in Table~\ref{tab:simulation_parameters}.

\subsubsection*{Isotropic turbulence}
The isotropic odor field is obtained by placing the odor source at the center of the simulation domain and forcing the flow field using a 2D isotropic, stochastic forcing term. The inference is operated on a two-dimensional horizontal slice at mid-height through the channel.

An earlier version of the code used for Computational Fluid Dynamics of odor transport is publicly available
at https://gitlab.com/vdv9265847/ IBbookVdV/, described in ref.~\cite{viola-book}. Our simulations were conducted using a GPU-accelerated version of the code that was developed by F.~Viola and colleagues in Refs.~\cite{viola2020fluid,viola2022fsei,viola2023high,viola-book} and will be shared publicly by these authors in the near future. All requests may be directed to F.~Viola and colleagues.

\begin{table} [h]

\setlength\tabcolsep{4pt}

\centering

\begin{tabular}{|c c c c c|} 
\hline
Simulation ID & $N_x$ & $N_y$ & $Re$  & $\eta/\Delta x$\\
\hline
(1) anisotropic large & 2000 & 500 & 17500   & 5.5\\
 (2) anisotropic small & 1024 & 256 & 7830  & 5\\
 (3) isotropic & 512 & 512 & 4990 & 0.8\\
 \hline

\end{tabular}
\caption{Parameters of the simulations: Simulation ID, number of grid points in the streamwise direction $N_x$, number of grid points in the cross-stream direction $N_y$, bulk Reynolds number $Re$ and ratio between Kolmogorov length $\eta$ and grid spacing $\Delta x$. The Reynolds number $Re=U_b (H/2) / \nu$ is based on the bulk speed $U_b$, half height $H/2$ and kinematic viscosity $\nu = \mu/\rho$ for the anisotropic simulations and $Re=U_\text{rms} L / \nu$, where $U_\text{rms}$ is the root mean square velocity in the midplane and $L$ is the length of the simulation domain, for the isotropic simulation.}
\label{tab:simulation_parameters}
\end{table}

\end{appendices}
\bibliography{references}

@article{TUTHILL2018R194,
title = {Proprioception},
journal = {Current Biology},
volume = {28},
number = {5},
pages = {R194-R203},
year = {2018},
issn = {0960-9822},
doi = {https://doi.org/10.1016/j.cub.2018.01.064},
url = {https://www.sciencedirect.com/science/article/pii/S0960982218300976},
author = {John C. Tuthill and Eiman Azim},

}

@article{bishop,
author = {Bishop, Chris M.},
title ={Training with noise is equivalent to Tikhonov regularization},
journal = {Neural computation},
volume = {7},
pages = {108--116},
year = {1995},
}

@article{stoch_resonance_revmodphys,
  title = {Stochastic resonance},
  author = {Gammaitoni, Luca and H\"anggi, Peter and Jung, Peter and Marchesoni, Fabio},
  journal = {Rev. Mod. Phys.},
  volume = {70},
  issue = {1},
  pages = {223--287},
  numpages = {0},
  year = {1998},
  month = {Jan},
  publisher = {American Physical Society},
  doi = {10.1103/RevModPhys.70.223},
  url = {https://link.aps.org/doi/10.1103/RevModPhys.70.223}
}

@article{benzi_sutera_vulpiani,
author = {Benzi, R. and Sutera, A. and Vulpiani, A.},
title ={The mechanism of stochastic resonance},
journal = {J. Phys. A, Math. Gen.},
volume = {14},
pages = {L453–L457},
year = {1981},
}

@article{adaptive_stoch_resonance,
author = {Mitaim, S. and Kosko, B.},
title ={Adaptive stochastic resonance},
journal = {Proc. IEEE},
volume = {86},
pages = {2152–2183},
year = {1998},
}

@article{KEATS2007465,
title = {Bayesian inference for source determination with applications to a complex urban environment},
journal = {Atmospheric Environment},
volume = {41},
number = {3},
pages = {465-479},
year = {2007},
issn = {1352-2310},
doi = {https://doi.org/10.1016/j.atmosenv.2006.08.044},
url = {https://www.sciencedirect.com/science/article/pii/S1352231006008703},
author = {Andrew Keats and Eugene Yee and Fue-Sang Lien},

}

@book{tarantola2005inverse,
  title={Inverse problem theory and methods for model parameter estimation},
  author={Tarantola, Albert},
  year={2005},
  publisher={SIAM}
}

@incollection{dashti2017bayesian,
  title={The Bayesian approach to inverse problems},
  author={Dashti, Masoumeh and Stuart, Andrew M},
  booktitle={Handbook of uncertainty quantification},
  pages={311--428},
  year={2017},
  publisher={Springer}
}

@article{Marcelo,
author = {Pereyra, Marcelo},
title = {Revisiting Maximum-A-Posteriori Estimation in Log-Concave Models},
journal = {SIAM Journal on Imaging Sciences},
volume = {12},
number = {1},
pages = {650-670},
year = {2019},
doi = {10.1137/18M1174076},

URL = { 
    
        https://doi.org/10.1137/18M1174076
    
    

},
eprint = { 
    
        https://doi.org/10.1137/18M1174076
    
    

}

}

@article{Clason,
author = {Clason, Christian and Helin, Tapio and Kretschmann, Remo and Piiroinen, Petteri},
title = {Generalized Modes in Bayesian Inverse Problems},
journal = {SIAM/ASA Journal on Uncertainty Quantification},
volume = {7},
number = {2},
pages = {652-684},
year = {2019},
doi = {10.1137/18M1191804},

URL = { 
    
        https://doi.org/10.1137/18M1191804
    
    

},
eprint = { 
    
        https://doi.org/10.1137/18M1191804
    
    

}

}

@article{Wacker2020MAPEF,
  title={MAP estimators for nonparametric Bayesian inverse problems in Banach spaces},
  author={Philipp Wacker},
  journal={arXiv: Probability},
  year={2020}
}

@article{Helin_2015,
doi = {10.1088/0266-5611/31/8/085009},
url = {https://dx.doi.org/10.1088/0266-5611/31/8/085009},
year = {2015},
month = {jul},
publisher = {IOP Publishing},
volume = {31},
number = {8},
pages = {085009},
author = {T Helin and M Burger},
title = {Maximum a posteriori probability estimates in infinite-dimensional Bayesian inverse problems},
journal = {Inverse Problems}

}

@article{Burger_2014,
doi = {10.1088/0266-5611/30/11/114004},
url = {https://dx.doi.org/10.1088/0266-5611/30/11/114004},
year = {2014},
month = {oct},
publisher = {IOP Publishing},
volume = {30},
number = {11},
pages = {114004},
author = {Martin Burger and Felix Lucka},
title = {Maximum a posteriori estimates in linear inverse problems with log-concave priors are proper Bayes estimators},
journal = {Inverse Problems}
}

@article{GRIBONVAL202149,
title = {On Bayesian estimation and proximity operators},
journal = {Applied and Computational Harmonic Analysis},
volume = {50},
pages = {49-72},
year = {2021},
issn = {1063-5203},
doi = {https://doi.org/10.1016/j.acha.2019.07.002},
url = {https://www.sciencedirect.com/science/article/pii/S1063520318301908},
author = {Rémi Gribonval and Mila Nikolova}

}

@article{bassett2019maximum,
  title={Maximum a posteriori estimators as a limit of Bayes estimators},
  author={Bassett, Robert and Deride, Julio},
  journal={Mathematical Programming},
  volume={174},
  pages={129--144},
  year={2019},
  publisher={Springer}
}

@article{DBLP:journals/corr/abs-1907-10902,
  author       = {Takuya Akiba and
                  Shotaro Sano and
                  Toshihiko Yanase and
                  Takeru Ohta and
                  Masanori Koyama},
  title        = {Optuna: {A} Next-generation Hyperparameter Optimization Framework},
  journal      = {CoRR},
  volume       = {abs/1907.10902},
  year         = {2019},
  url          = {http://arxiv.org/abs/1907.10902},
  eprinttype    = {arXiv},
  eprint       = {1907.10902},
  bibsource    = {dblp computer science bibliography, https://dblp.org}
}

@article{baker2018algorithms,
  title={Algorithms for olfactory search across species},
  author={Baker, Keeley L and Dickinson, Michael and Findley, Teresa M and Gire, David H and Louis, Matthieu and Suver, Marie P and Verhagen, Justus V and Nagel, Katherine I and Smear, Matthew C},
  journal={Journal of Neuroscience},
  volume={38},
  number={44},
  pages={9383--9389},
  year={2018},
  publisher={Soc Neuroscience}
}

@article{murlis1992odor,
  title={Odor plumes and how insects use them},
  author={Murlis, John and Elkinton, Joseph S and Carde, Ring T},
  journal={Annual review of entomology},
  volume={37},
  number={1},
  pages={505--532},
  year={1992},
  publisher={Annual Reviews 4139 El Camino Way, PO Box 10139, Palo Alto, CA 94303-0139, USA}
}

@article{durve2020collective,
  title={Collective olfactory search in a turbulent environment},
  author={Durve, Mihir and Piro, Lorenzo and Cencini, Massimo and Biferale, Luca and Celani, Antonio},
  journal={Physical Review E},
  volume={102},
  number={1},
  pages={012402},
  year={2020},
  publisher={APS}
}

@article{maselli2020sensorial,
  title={Sensorial hierarchy in Octopus vulgaris’s food choice: Chemical vs. visual},
  author={Maselli, Valeria and Al-Soudy, Al-Sayed and Buglione, Maria and Aria, Massimo and Polese, Gianluca and Di Cosmo, Anna},
  journal={Animals},
  volume={10},
  number={3},
  pages={457},
  year={2020},
  publisher={MDPI}
}

@article{van2020molecular,
  title={Molecular basis of chemotactile sensation in octopus},
  author={Van Giesen, Lena and Kilian, Peter B and Allard, Corey AH and Bellono, Nicholas W},
  journal={Cell},
  volume={183},
  number={3},
  pages={594--604},
  year={2020},
  publisher={Elsevier}
}

@book{wells2013octopus,
  title={Octopus: physiology and behaviour of an advanced invertebrate},
  author={Wells, Martin John},
  year={2013},
  publisher={Springer Science \& Business Media}
}

@article{villanueva2017cephalopods,
  title={Cephalopods as predators: a short journey among behavioral flexibilities, adaptions, and feeding habits},
  author={Villanueva, Roger and Perricone, Valentina and Fiorito, Graziano},
  journal={Frontiers in Physiology},
  volume={8},
  pages={598},
  year={2017},
  publisher={Frontiers Media SA}
}

@article{al2021identification,
  title={Identification and characterization of a rhodopsin kinase gene in the suckers of Octopus vulgaris: Looking around using arms?},
  author={Al-Soudy, Al-Sayed and Maselli, Valeria and Galdiero, Stefania and Kuba, Michael J and Polese, Gianluca and Di Cosmo, Anna},
  journal={Biology},
  volume={10},
  number={9},
  pages={936},
  year={2021},
  publisher={MDPI}
}

@article{allard2023structural,
  title={Structural basis of sensory receptor evolution in octopus},
  author={Allard, Corey AH and Kang, Guipeun and Kim, Jeong Joo and Valencia-Montoya, Wendy A and Hibbs, Ryan E and Bellono, Nicholas W},
  journal={Nature},
  volume={616},
  number={7956},
  pages={373--377},
  year={2023},
  publisher={Nature Publishing Group UK London}
}

@article{kang2023sensory,
  title={Sensory specializations drive octopus and squid behaviour},
  author={Kang, Guipeun and Allard, Corey AH and Valencia-Montoya, Wendy A and van Giesen, Lena and Kim, Jeong Joo and Kilian, Peter B and Bai, Xiaochen and Bellono, Nicholas W and Hibbs, Ryan E},
  journal={Nature},
  volume={616},
  number={7956},
  pages={378--383},
  year={2023},
  publisher={Nature Publishing Group UK London}
}

@article{gutnick2011octopus,
  title={Octopus vulgaris uses visual information to determine the location of its arm},
  author={Gutnick, Tamar and Byrne, Ruth A and Hochner, Binyamin and Kuba, Michael},
  journal={Current biology},
  volume={21},
  number={6},
  pages={460--462},
  year={2011},
  publisher={Elsevier}
}

@article{gutnick2020use,
  title={Use of peripheral sensory information for central nervous control of arm movement by Octopus vulgaris},
  author={Gutnick, Tamar and Zullo, Letizia and Hochner, Binyamin and Kuba, Michael J},
  journal={Current Biology},
  volume={30},
  number={21},
  pages={4322--4327},
  year={2020},
  publisher={Elsevier}
}

@article{sumbre2005motor,
  title={Motor control of flexible octopus arms},
  author={Sumbre, Germ{\'a}n and Fiorito, Graziano and Flash, Tamar and Hochner, Binyamin},
  journal={Nature},
  volume={433},
  number={7026},
  pages={595--596},
  year={2005},
  publisher={Nature Publishing Group UK London}
}

@article{orsi2021scalar,
  title={Scalar mixing in homogeneous isotropic turbulence: A numerical study},
  author={Orsi, Michel and Soulhac, Lionel and Feraco, Fabio and Marro, Massimo and Rosenberg, Duane and Marino, Raffaele and Boffadossi, Maurizio and Salizzoni, Pietro},
  journal={Physical Review Fluids},
  volume={6},
  number={3},
  pages={034502},
  year={2021},
  publisher={APS}
}

@article{villermaux2003mixing,
  title={Mixing is an aggregation process},
  author={Villermaux, Emmanuel and Duplat, J{\'e}r{\^o}me},
  journal={Comptes Rendus M{\'e}canique},
  volume={331},
  number={7},
  pages={515--523},
  year={2003},
  publisher={Elsevier}
}

@article{duplat2008mixing,
  title={Mixing by random stirring in confined mixtures},
  author={Duplat, Jer{\^o}me and Villermaux, Emmanuel},
  journal={Journal of Fluid Mechanics},
  volume={617},
  pages={51--86},
  year={2008},
  publisher={Cambridge University Press}
}

@article{shraiman2000scalar,
  title={Scalar turbulence},
  author={Shraiman, Boris I and Siggia, Eric D},
  journal={Nature},
  volume={405},
  number={6787},
  pages={639--646},
  year={2000},
  publisher={Nature Publishing Group UK London}
}

@article{celani2014odor,
  title={Odor landscapes in turbulent environments},
  author={Celani, Antonio and Villermaux, Emmanuel and Vergassola, Massimo},
  journal={Physical Review X},
  volume={4},
  number={4},
  pages={041015},
  year={2014},
  publisher={APS}
}

@article{rigolli2022alternation,
  title={Alternation emerges as a multi-modal strategy for turbulent odor navigation},
  author={Rigolli, Nicola and Reddy, Gautam and Seminara, Agnese and Vergassola, Massimo},
  journal={Elife},
  volume={11},
  year={2022},
  publisher={eLife Sciences Publications, Ltd}
}

@article{rigolli_learning_2022,
	title = {Learning to predict target location with turbulent odor plumes},
	volume = {11},
	issn = {2050-084X},
	url = {https://elifesciences.org/articles/72196},
	doi = {10.7554/eLife.72196},
	language = {en},
	urldate = {2024-04-10},
	journal = {eLife},
	author = {Rigolli, Nicola and Magnoli, Nicodemo and Rosasco, Lorenzo and Seminara, Agnese},
	month = aug,
	year = {2022},
	pages = {e72196},
}

@article{victor,
	title = {Olfactory navigation and the receptor nonlinearity.},
	volume = {39},
	number = {7},
	urldate = {2024-04-10},
	journal = {The Journal of Neuroscience},
	author = {Victor, Jonathan D. and Boie, Sebastian D. and Connor, Erin G. and Crimaldi, John P. and Ermentrout, G. and  Nagel, Katherine I.},
	year = {2019},
	pages = {3713 -–3727},
}

@article{boie_information-theoretic_2018,
	title = {Information-theoretic analysis of realistic odor plumes: {What} cues are useful for determining location?},
	volume = {14},
	issn = {1553-7358},
	shorttitle = {Information-theoretic analysis of realistic odor plumes},
	url = {https://dx.plos.org/10.1371/journal.pcbi.1006275},
	doi = {10.1371/journal.pcbi.1006275},
	language = {en},
	number = {7},
	urldate = {2024-04-10},
	journal = {PLOS Computational Biology},
	author = {Boie, Sebastian D. and Connor, Erin G. and McHugh, Margaret and Nagel, Katherine I. and Ermentrout, G. Bard and Crimaldi, John P. and Victor, Jonathan D.},
	editor = {Louis, Matthieu},
	month = jul,
	year = {2018},
	pages = {e1006275},
}

@article{CRLB_noise,
author = {Balkan, Gökce Osman and Sinan, Gezici},
title = {CRLB based optimal noise enhanced parameter estimation using quantized observations},
volume = {17},
journal = {IEEE Signal Processing Letters },
pages = {477 -- 480},
year = {2010}
}

@article{bayesian_misspecified,
author = {Robert H. Berk},
title = {{Limiting Behavior of Posterior Distributions when the Model is Incorrect}},
volume = {37},
journal = {The Annals of Mathematical Statistics},
number = {1},
publisher = {Institute of Mathematical Statistics},
pages = {51 -- 58},
year = {1966},
doi = {10.1214/aoms/1177699597},
URL = {https://doi.org/10.1214/aoms/1177699597}
}

@article{bayesian_convergence,
author="LE CAM L.",
title="On some asymptotic properties of maximum likelihood estimates and related Bayes' estimates",
journal="Univ. Calif. Publ. in Statist.",
year="1953",
volume="1",
pages="277-330",
URL="https://cir.nii.ac.jp/crid/1572261550097442816"
}

@article{celani_odor_2014,
	title = {Odor {Landscapes} in {Turbulent} {Environments}},
	volume = {4},
	copyright = {http://creativecommons.org/licenses/by/3.0/},
	issn = {2160-3308},
	url = {https://link.aps.org/doi/10.1103/PhysRevX.4.041015},
	doi = {10.1103/PhysRevX.4.041015},
	language = {en},
	number = {4},
	urldate = {2024-04-10},
	journal = {Physical Review X},
	author = {Celani, Antonio and Villermaux, Emmanuel and Vergassola, Massimo},
	month = oct,
	year = {2014},
	pages = {041015},
}

@misc{piro_many_2024,
	title = {Many wrong models approach to localize an odor source in turbulence: introducing the weighted {Bayesian} update},
	shorttitle = {Many wrong models approach to localize an odor source in turbulence},
	url = {http://arxiv.org/abs/2407.08343},
	language = {en},
	urldate = {2024-07-17},
	publisher = {arXiv},
	author = {Piro, Lorenzo and Heinonen, Robin A. and Cencini, Massimo and Biferale, Luca},
	month = jul,
	year = {2024},
	note = {arXiv:2407.08343 [physics]},
}

@article{heinonen2025exploring,
  title={Exploring Bayesian olfactory search in realistic turbulent flows},
  author={Heinonen, Robin A and Biferale, Luca and Celani, Antonio and Vergassola, Massimo},
  journal={Physical Review Fluids},
  volume={10},
  number={6},
  pages={064614},
  year={2025},
  publisher={APS}
}

@book{carroll2006measurement,
  title={Measurement error in nonlinear models: a modern perspective},
  author={Carroll, Raymond J and Ruppert, David and Stefanski, Leonard A and Crainiceanu, Ciprian M},
  year={2006},
  publisher={Chapman and Hall/CRC}
}

@article{mcdonnell2009stochastic,
  title={What is stochastic resonance? Definitions, misconceptions, debates, and its relevance to biology},
  author={McDonnell, Mark D and Abbott, Derek},
  journal={PLoS computational biology},
  volume={5},
  number={5},
  pages={e1000348},
  year={2009},
  publisher={Public Library of Science San Francisco, USA}
}

@article {10.7554/eLife.102906,
article_type = {journal},
title = {Q-learning with temporal memory to navigate turbulence},
author = {Rando, Marco and James, Martin and Verri, Alessandro and Rosasco, Lorenzo and Seminara, Agnese},
editor = {Berman, Gordon J and Walczak, Aleksandra M},
volume = 13,
year = 2025,
month = {jul},
pub_date = {2025-07-21},
pages = {RP102906},
citation = {eLife 2025;13:RP102906},
doi = {10.7554/eLife.102906},
url = {https://doi.org/10.7554/eLife.102906},
journal = {eLife},
issn = {2050-084X},
publisher = {eLife Sciences Publications, Ltd},
}

@article{hochner2023embodied,
  title={Embodied mechanisms of motor control in the octopus},
  author={Hochner, Binyamin and Zullo, Letizia and Shomrat, Tal and Levy, Guy and Nesher, Nir},
  journal={Current Biology},
  volume={33},
  number={20},
  pages={R1119--R1125},
  year={2023},
}

@article{zullo2011new,
  title={A new perspective on the organization of an invertebrate brain},
  author={Zullo, Letizia and Hochner, Binyamin},
  journal={Communicative \& integrative biology},
  volume={4},
  number={1},
  pages={26--29},
  year={2011},
  publisher={Taylor \& Francis}
}

@article{nesher2014self,
  title={Self-recognition mechanism between skin and suckers prevents octopus arms from interfering with each other},
  author={Nesher, Nir and Levy, Guy and Grasso, Frank W and Hochner, Binyamin},
  journal={Current biology},
  volume={24},
  number={11},
  pages={1271--1275},
  year={2014},
  publisher={Elsevier}
}

@article{sivitilli2023mechanisms,
  title={Mechanisms of octopus arm search behavior without visual feedback},
  author={Sivitilli, Dominic M and Strong, Terrell and Weertman, Willem and Ullmann, Joseph and Smith, Joshua R and Gire, David H},
  journal={Bioinspiration \& Biomimetics},
  volume={18},
  number={6},
  pages={066017},
  year={2023},
  publisher={IOP Publishing}
}

@article{hochner2013nervous,
  title={How nervous systems evolve in relation to their embodiment: what we can learn from octopuses and other molluscs},
  author={Hochner, Binyamin},
  journal={Brain Behavior and Evolution},
  volume={82},
  number={1},
  pages={19--30},
  year={2013},
  publisher={S. Karger AG}
}

@article{tarvin2020sucker,
  title={A Sucker for Taste},
  author={Tarvin, Rebecca D},
  journal={Cell},
  volume={183},
  number={3},
  pages={587--588},
  year={2020},
  publisher={Elsevier}
}

@article{gutfreund1998,
  title={Patterns of arm muscle activation involved in octopus reaching movements},
  author={Gutfreund, Yoram and Flash, Tamar and Fiorito, Graziano and Hochner, Binyamin},
  journal={Journal of Neuroscience},
  volume={18},
  number={15},
  pages={5976--5987},
  year={1998},
  publisher={Society for Neuroscience}
}

@article{gutfreund1996,
  title={Organization of octopus arm movements: a model system for studying the control of flexible arms},
  author={Gutfreund, Yoram and Flash, Tamar and Yarom, Yosef and Fiorito, Graziano and Segev, Idan and Hochner, Binyamin},
  journal={Journal of Neuroscience},
  volume={16},
  number={22},
  pages={7297--7307},
  year={1996},
  publisher={Society for Neuroscience}
}

@article{zullo2025,
  title={Proprioception in muscle hydrostats},
  author={Zullo, Letizia and R{\"o}ckner, Janina Leonie and Pistolato, Beatrice},
  journal={Integrative And Comparative Biology},
  pages={icaf046},
  year={2025},
  publisher={Oxford University Press}
}

@article{kowadlo2008robot,
  title={Robot odor localization: a taxonomy and survey},
  author={Kowadlo, Gideon and Russell, R Andrew},
  journal={The International Journal of Robotics Research},
  volume={27},
  number={8},
  pages={869--894},
  year={2008},
  publisher={SAGE Publications Sage UK: London, England}
}

@article{nehorai1995detection,
  title={Detection and localization of vapor-emitting sources},
  author={Nehorai, Arye and Porat, Boaz and Paldi, Eytan},
  journal={IEEE Transactions on Signal Processing},
  volume={43},
  number={1},
  pages={243--253},
  year={1995},
  publisher={IEEE}
}

@article{alpay2002model,
  title={Model-based solution techniques for the source localization problem},
  author={Alpay, Mehmet E and Shor, Molly H},
  journal={IEEE transactions on control systems technology},
  volume={8},
  number={6},
  pages={895--904},
  year={2002},
  publisher={IEEE}
}

@article{wang2018toward,
  title={Toward perceptive soft robots: Progress and challenges},
  author={Wang, Hongbo and Totaro, Massimo and Beccai, Lucia},
  journal={Advanced science},
  volume={5},
  number={9},
  pages={1800541},
  year={2018},
  publisher={Wiley Online Library}
}

@article{viola2023high,
  title={High-fidelity model of the human heart: an immersed boundary implementation},
  author={Viola, Francesco and Del Corso, Giulio and Verzicco, Roberto},
  journal={Physical Review Fluids},
  volume={8},
  number={10},
  pages={100502},
  year={2023},
  publisher={APS}
}

@article{viola2022fsei,
  title={FSEI-GPU: GPU accelerated simulations of the fluid--structure--electrophysiology interaction in the left heart},
  author={Viola, Francesco and Spandan, Vamsi and Meschini, Valentina and Romero, Joshua and Fatica, Massimiliano and de Tullio, Marco D and Verzicco, Roberto},
  journal={Computer physics communications},
  volume={273},
  pages={108248},
  year={2022},
  publisher={Elsevier}
}

@article{viola2020fluid,
  title={Fluid--structure-electrophysiology interaction (FSEI) in the left-heart: a multi-way coupled computational model},
  author={Viola, Francesco and Meschini, Valentina and Verzicco, Roberto},
  journal={European Journal of Mechanics-B/Fluids},
  volume={79},
  pages={212--232},
  year={2020},
  publisher={Elsevier}
}

@book{viola-book, 
place={Cambridge}, 
series={Cambridge Monographs on Applied and Computational Mathematics}, 
title={An Introduction to Immersed Boundary Methods}, 
publisher={Cambridge University Press}, 
author={Verzicco, Roberto and de Tullio, Marco D. and Viola, Francesco}, 
year={2025}, 
collection={Cambridge Monographs on Applied and Computational Mathematics}}

@article{xie2023octopus,
  title={Octopus-inspired sensorized soft arm for environmental interaction},
  author={Xie, Zhexin and Yuan, Feiyang and Liu, Jiaqi and Tian, Lufeng and Chen, Bohan and Fu, Zhongqiang and Mao, Sizhe and Jin, Tongtong and Wang, Yun and He, Xia and others},
  journal={Science Robotics},
  volume={8},
  number={84},
  pages={eadh7852},
  year={2023},
  publisher={American Association for the Advancement of Science}
}

@article{giordano2021perspective,
  title={A perspective on cephalopods mimicry and bioinspired technologies toward proprioceptive autonomous soft robots},
  author={Giordano, Goffredo and Carlotti, Marco and Mazzolai, Barbara},
  journal={Advanced Materials Technologies},
  volume={6},
  number={12},
  pages={2100437},
  year={2021},
  publisher={Wiley Online Library}
}

@inproceedings{russell1995robotic,
  title={A robotic system to locate hazardous chemical leaks},
  author={Russell, R Andrew and Thiel, David and Deveza, Reimundo and Mackay-Sim, Alan},
  booktitle={Proceedings of 1995 IEEE international conference on robotics and automation},
  volume={1},
  pages={556--561},
  year={1995},
  organization={IEEE}
}

@article{truby2020distributed,
  title={Distributed proprioception of 3D configuration in soft, sensorized robots via deep learning},
  author={Truby, Ryan L and Della Santina, Cosimo and Rus, Daniela},
  journal={IEEE Robotics and Automation Letters},
  volume={5},
  number={2},
  pages={3299--3306},
  year={2020},
  publisher={IEEE}
}

@article{kim2021review,
  title={Review of machine learning methods in soft robotics},
  author={Kim, Daekyum and Kim, Sang-Hun and Kim, Taekyoung and Kang, Brian Byunghyun and Lee, Minhyuk and Park, Wookeun and Ku, Subyeong and Kim, DongWook and Kwon, Junghan and Lee, Hochang and others},
  journal={Plos one},
  volume={16},
  number={2},
  pages={e0246102},
  year={2021},
  publisher={Public Library of Science San Francisco, CA USA}
}

@article{matthes2005source,
  title={Source localization by spatially distributed electronic noses for advection and diffusion},
  author={Matthes, J{\"o}rg and Groll, Lutz and Keller, Hubert B},
  journal={IEEE Transactions on Signal Processing},
  volume={53},
  number={5},
  pages={1711--1719},
  year={2005},
  publisher={IEEE}
}

@article{chen2019odor,
  title={Odor source localization algorithms on mobile robots: A review and future outlook},
  author={Chen, Xin-xing and Huang, Jian},
  journal={Robotics and Autonomous Systems},
  volume={112},
  pages={123--136},
  year={2019},
  publisher={Elsevier}
}

@article{tytell,
    author = {Tytell, Eric},
    title = {IS THERE AN OCTOPUNCULUS?},
    journal = {Journal of Experimental Biology},
    volume = {213},
    number = {4},
    pages = {v-v},
    year = {2010},
    month = {02},
    issn = {0022-0949},
    doi = {10.1242/jeb.036376},
    url = {https://doi.org/10.1242/jeb.036376},
    eprint = {https://journals.biologists.com/jeb/article-pdf/213/4/v/1271873/osiv-vii.pdf},
}

@article{zullo2009nonsomatotopic,
  title={Nonsomatotopic organization of the higher motor centers in octopus},
  author={Zullo, Letizia and Sumbre, German and Agnisola, Claudio and Flash, Tamar and Hochner, Binyamin},
  journal={Current Biology},
  volume={19},
  number={19},
  pages={1632--1636},
  year={2009},
  publisher={Elsevier}
}

@article{weertman2024octopus,
  title={Octopus can use odor plumes to find food},
  author={Weertman, Willem Lee and Gopal, Venkatesh and Sivitilli, Dominic M and Scheel, David and Gire, David H},
  journal={bioRxiv},
  pages={2024--08},
  year={2024},
  publisher={Cold Spring Harbor Laboratory}
}

@article{noda2019bayesian,
  title={Bayesian Inference from Symplectic Geometric Viewpoint},
  author={Noda, Tomonori and Matsuyama, Hinako},
  journal={Advances in Pure Mathematics},
  volume={9},
  pages={827--831},
  year={2019},
  publisher={Scientific Research Publishing, Inc.}
}

@article{mamiya2018neural,
  title={Neural coding of leg proprioception in Drosophila},
  author={Mamiya, Akira and Gurung, Pralaksha and Tuthill, John C},
  journal={Neuron},
  volume={100},
  number={3},
  pages={636--650},
  year={2018},
  publisher={Elsevier}
}

@article{oliver2021molecular,
  title={Molecular correlates of muscle spindle and Golgi tendon organ afferents},
  author={Oliver, Katherine M and Florez-Paz, Danny M and Badea, Tudor Constantin and Mentis, George Z and Menon, Vilas and de Nooij, Joriene C},
  journal={Nature communications},
  volume={12},
  number={1},
  pages={1451},
  year={2021},
  publisher={Nature Publishing Group UK London}
}

@article{flash2023biomechanics,
  title={Biomechanics, motor control and dynamic models of the soft limbs of the octopus and other cephalopods},
  author={Flash, Tamar and Zullo, Letizia},
  journal={Journal of Experimental Biology},
  volume={226},
  number={Suppl\_1},
  pages={jeb245295},
  year={2023},
  publisher={The Company of Biologists Ltd}
}

@article{crook2014neuroethology,
  title={Neuroethology: self-recognition helps octopuses avoid entanglement},
  author={Crook, Robyn J and Walters, Edgar T},
  journal={Current biology},
  volume={24},
  number={11},
  pages={R520--R521},
  year={2014},
  publisher={Elsevier}
}

@article{hubel1962receptive,
  title={Receptive fields, binocular interaction and functional architecture in the cat's visual cortex},
  author={Hubel, David H and Wiesel, Torsten N},
  journal={The Journal of physiology},
  volume={160},
  number={1},
  pages={106},
  year={1962}
}

@article{hubel1959receptive,
  title={Receptive fields of single neurones in the cat's striate cortex},
  author={Hubel, David H and Wiesel, Torsten N},
  journal={The Journal of physiology},
  volume={148},
  number={3},
  pages={574},
  year={1959}
}

@article{decharms1996primary,
  title={Primary cortical representation of sounds by the coordination of action-potential timing},
  author={Decharms, R Christopher and Merzenich, Michael M},
  journal={Nature},
  volume={381},
  number={6583},
  pages={610--613},
  year={1996},
  publisher={Nature Publishing Group UK London}
}

@article{romani1982tonotopic,
  title={Tonotopic organization of the human auditory cortex},
  author={Romani, Gian Luca and Williamson, Samuel J and Kaufman, Lloyd},
  journal={Science},
  volume={216},
  number={4552},
  pages={1339--1340},
  year={1982},
  publisher={American Association for the Advancement of Science}
}

@article{woolsey1970structural,
  title={The structural organization of layer IV in the somatosensory region (SI) of mouse cerebral cortex: the description of a cortical field composed of discrete cytoarchitectonic units},
  author={Woolsey, Thomas A and Van der Loos, Hendrik},
  journal={Brain research},
  volume={17},
  number={2},
  pages={205--242},
  year={1970},
  publisher={Elsevier}
}

@article{petersen2007functional,
  title={The functional organization of the barrel cortex},
  author={Petersen, Carl CH},
  journal={Neuron},
  volume={56},
  number={2},
  pages={339--355},
  year={2007},
  publisher={Elsevier}
}

@article{petersen2012attention,
  title={The attention system of the human brain: 20 years after},
  author={Petersen, Steven E and Posner, Michael I},
  journal={Annual review of neuroscience},
  volume={35},
  number={1},
  pages={73--89},
  year={2012},
  publisher={Annual Reviews}
}

@article{ma2006bayesian,
  title={Bayesian inference with probabilistic population codes},
  author={Ma, Wei Ji and Beck, Jeffrey M and Latham, Peter E and Pouget, Alexandre},
  journal={Nature neuroscience},
  volume={9},
  number={11},
  pages={1432--1438},
  year={2006},
  publisher={Nature Publishing Group US New York}
}

@article{li2017neural,
  title={Neural code—neural self-information theory on how cell-assembly code rises from spike time and neuronal variability},
  author={Li, Meng and Tsien, Joe Z},
  journal={Frontiers in cellular neuroscience},
  volume={11},
  pages={236},
  year={2017},
  publisher={Frontiers Media SA}
}

@article{ermentrout2008reliability,
  title={Reliability, synchrony and noise},
  author={Ermentrout, G Bard and Gal{\'a}n, Roberto F and Urban, Nathaniel N},
  journal={Trends in neurosciences},
  volume={31},
  number={8},
  pages={428--434},
  year={2008},
  publisher={Elsevier}
}

@article{knill2004bayesian,
  title={The Bayesian brain: the role of uncertainty in neural coding and computation},
  author={Knill, David C and Pouget, Alexandre},
  journal={TRENDS in Neurosciences},
  volume={27},
  number={12},
  pages={712--719},
  year={2004},
  publisher={Elsevier}
}

@article{moss2004stochastic,
  title={Stochastic resonance and sensory information processing: a tutorial and review of application},
  author={Moss, Frank and Ward, Lawrence M and Sannita, Walter G},
  journal={Clinical neurophysiology},
  volume={115},
  number={2},
  pages={267--281},
  year={2004},
  publisher={Elsevier}
}

\clearpage
\pagebreak

\widetext
\begin{center}
\textbf{\large Supporting Information}
\end{center}

\setcounter{equation}{0}
\setcounter{figure}{0}
\setcounter{table}{0}
\setcounter{page}{1}
\setcounter{section}{0}
\makeatletter
\renewcommand{\theequation}{S\arabic{equation}}
\renewcommand{\thefigure}{S\arabic{figure}}
\renewcommand{\thetable}{S\arabic{table}}
\begin{figure}[h]
\includegraphics[width=\linewidth]{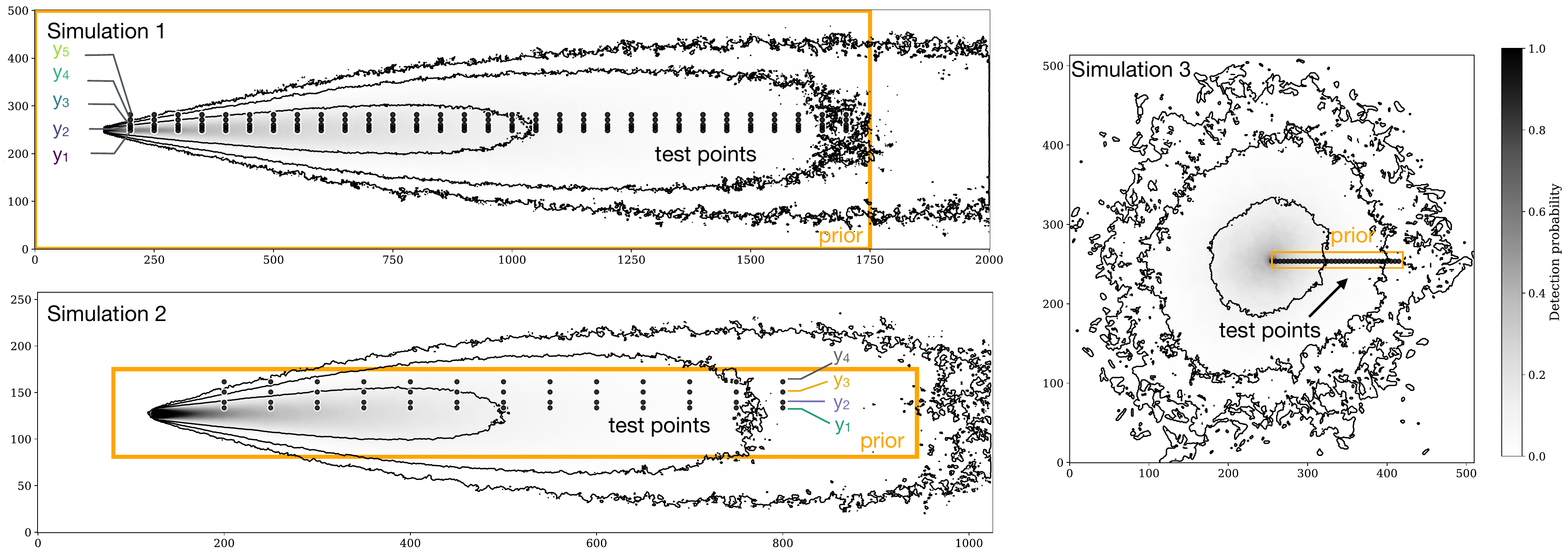} 
\caption{\label{fig:S1}
Test locations for simulation 1, 2 and 3. 
Simulation 1: $y=0,5,10,15,20$; prior: $1750\times 500$. Simulation 2: $y=0,5,10,15,20$; prior: $863\times 94$. Simulation 3: test points positioned radially in 1D; length of prior: 165. In all panels, increasingly wide contours of the likelihood (black) corresponding to $\ell=0.1,\,0.01$, and $0.001$ and all priors are flat within the region indicated in orange.}
\end{figure}

\begin{figure}
\includegraphics[width=0.75\linewidth]{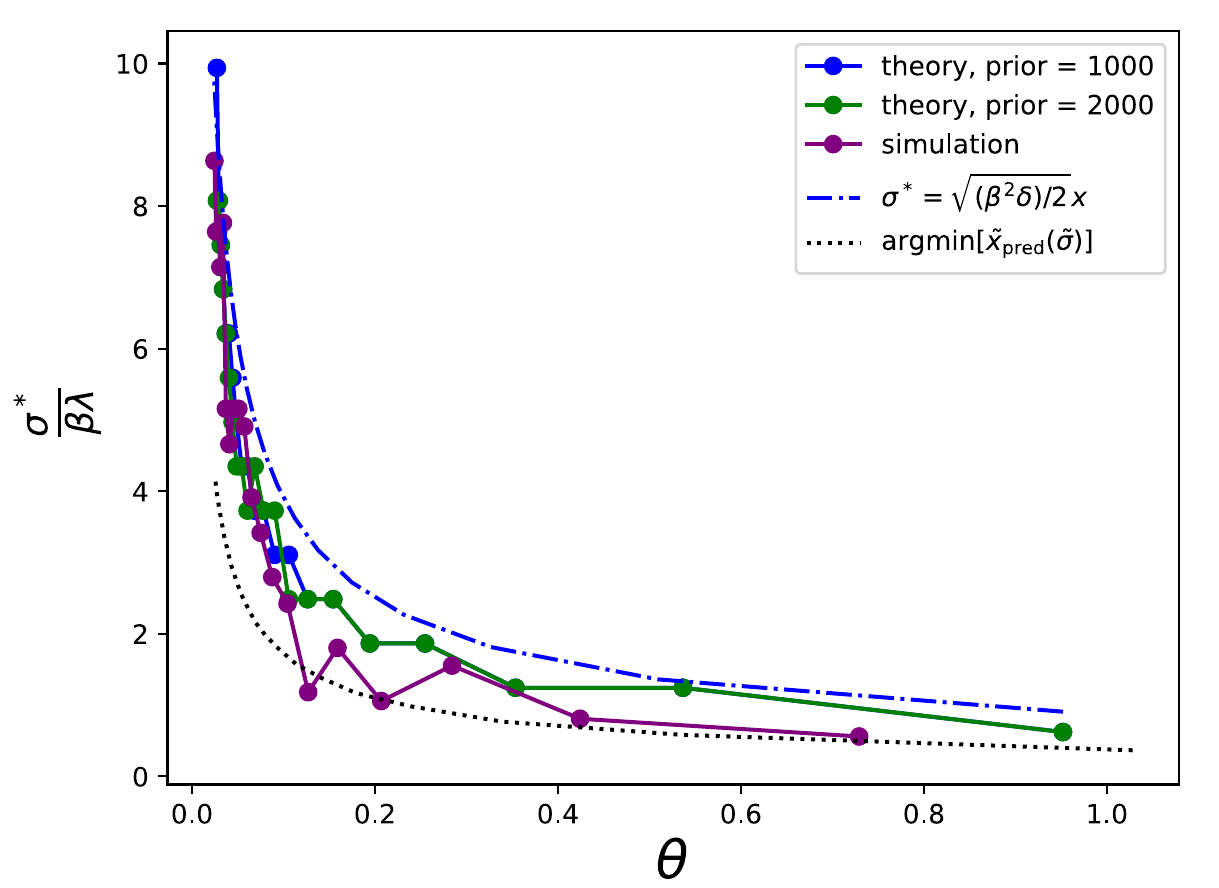} 
\caption{\label{fig:S2} Theoretically optimal size of the multisensor agent, $\sigma^*$, normalized with the parameters obtained by fitting the empirical likelihood with the analytic form (eq.~8 main text), as a function of the fraction of detections in the $N\rightarrow \infty$ limit, where $\theta = \ell(x,0)$.
}
\end{figure}

\begin{figure}
\includegraphics[width=0.8\linewidth]{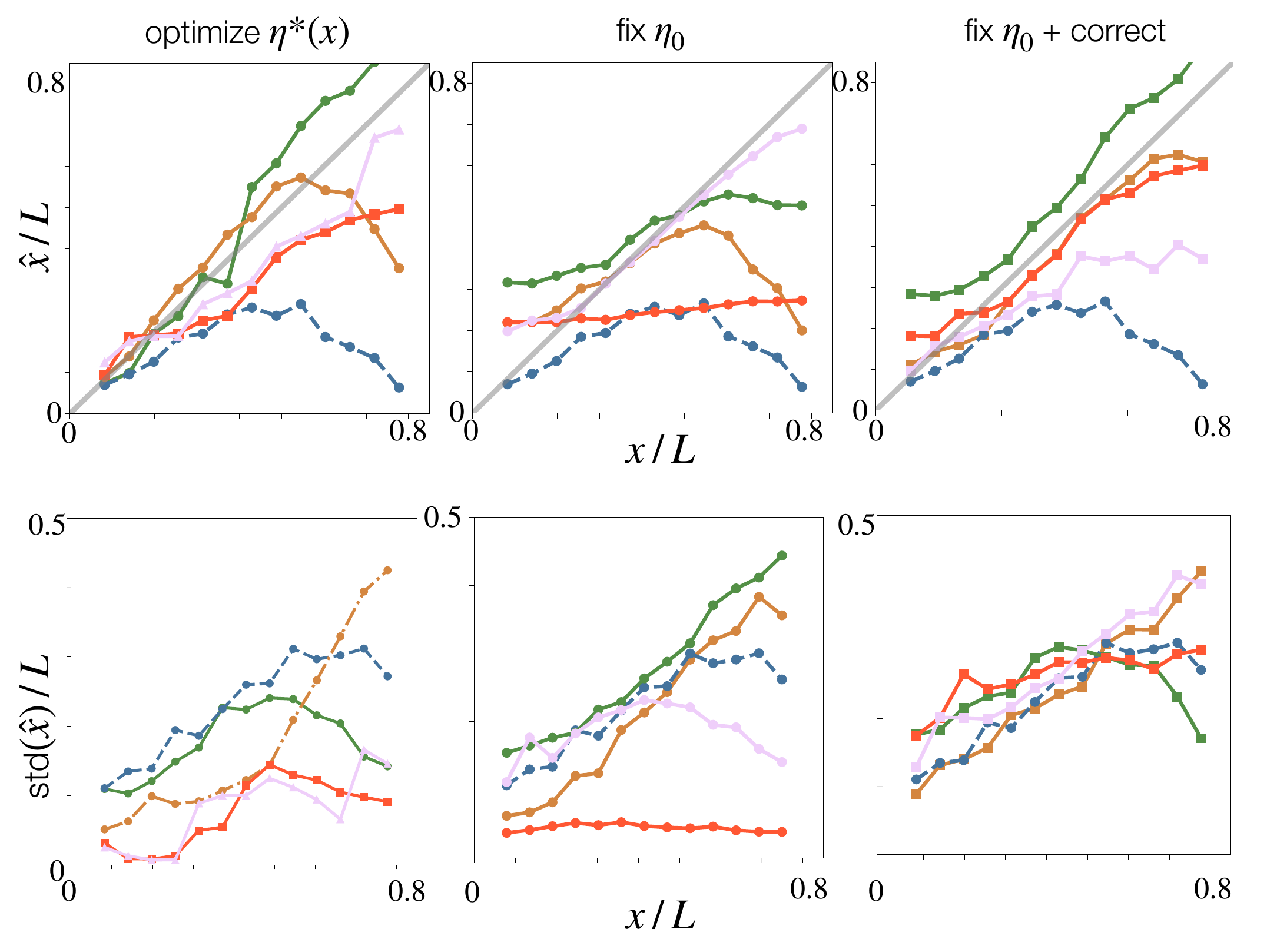} 
\caption{\label{fig:S3} Estimated position $\hat{x}$ (top) and standard deviation std$(\hat{x})$ (bottom) of the source \emph{vs} ground truth $x$. All quantities are normalized with the length of the prior, $L$. Different colors correspond to different sources of noise (Color code as in Fig.~S4; blue = perfect proprioception) and different columns correspond to different ways of tuning the error: left, noise is tuned to its optimal value; center: noise is fixed; right: noise is fixed and error correction is applied to the inference.}
\end{figure}

\begin{figure}
\includegraphics[width=1\linewidth]{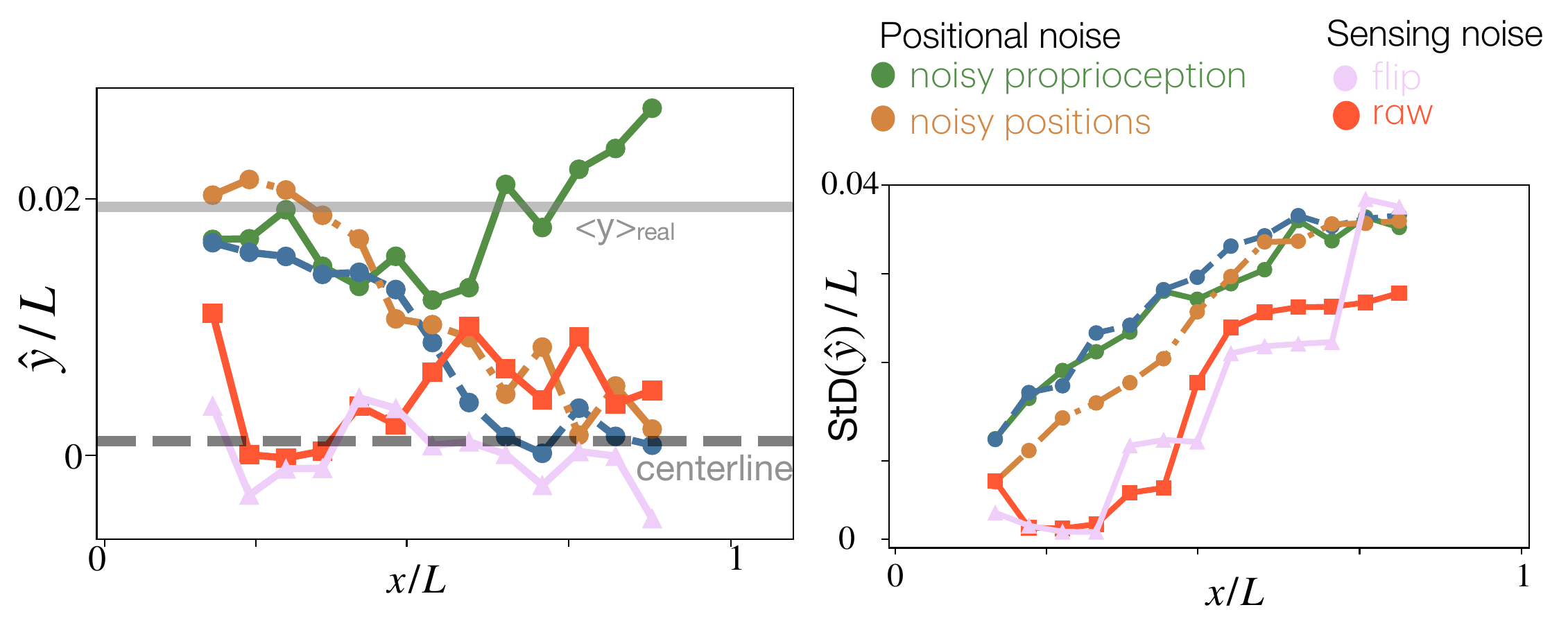} 
\caption{\label{fig:S4}
Estimated position $\hat{y}$ (left) and standard deviation std$(\hat{y})$ (right) of the source \emph{vs} ground truth $x$. For simplicity, we averaged the $y$ prediction across all four values. The gray line represents the correct prediction of the average $y$, and is marked with $<y>_{real} = (y_1+y_2+y_3+y_4)/4$. All quantities are normalized with the length of the prior, $L$. Different colors correspond to different sources of noise according to the legend, with noise tuned to the optimal value, defined as the one that maximizes accuracy in the $x$ direction. For positional noise, optimizing inference in $x$ still allows to infer $y$  similar to perfect proprioception (blue curve). Sensing noise, optimized for inference in $x$, precludes inference in $y$.}
\end{figure}

\begin{figure}
\includegraphics[width=1\linewidth]{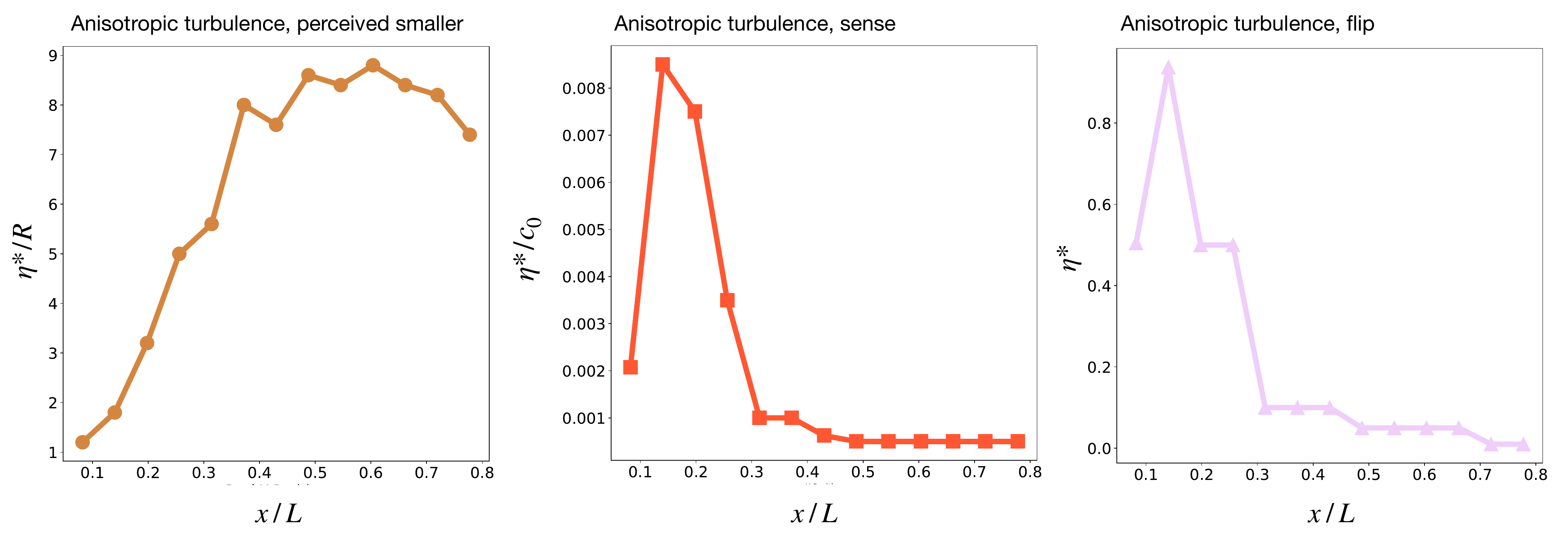} 
\caption{\label{fig:S5}
Normalized optimal errors, as a function of distance of the test point from the origin, for positional error on the actual sensor position (left), sensing error affecting the concentration (center) and flipping error, affecting the probability to detect (right).}
\end{figure}

\begin{figure}
\includegraphics[width=1\linewidth]{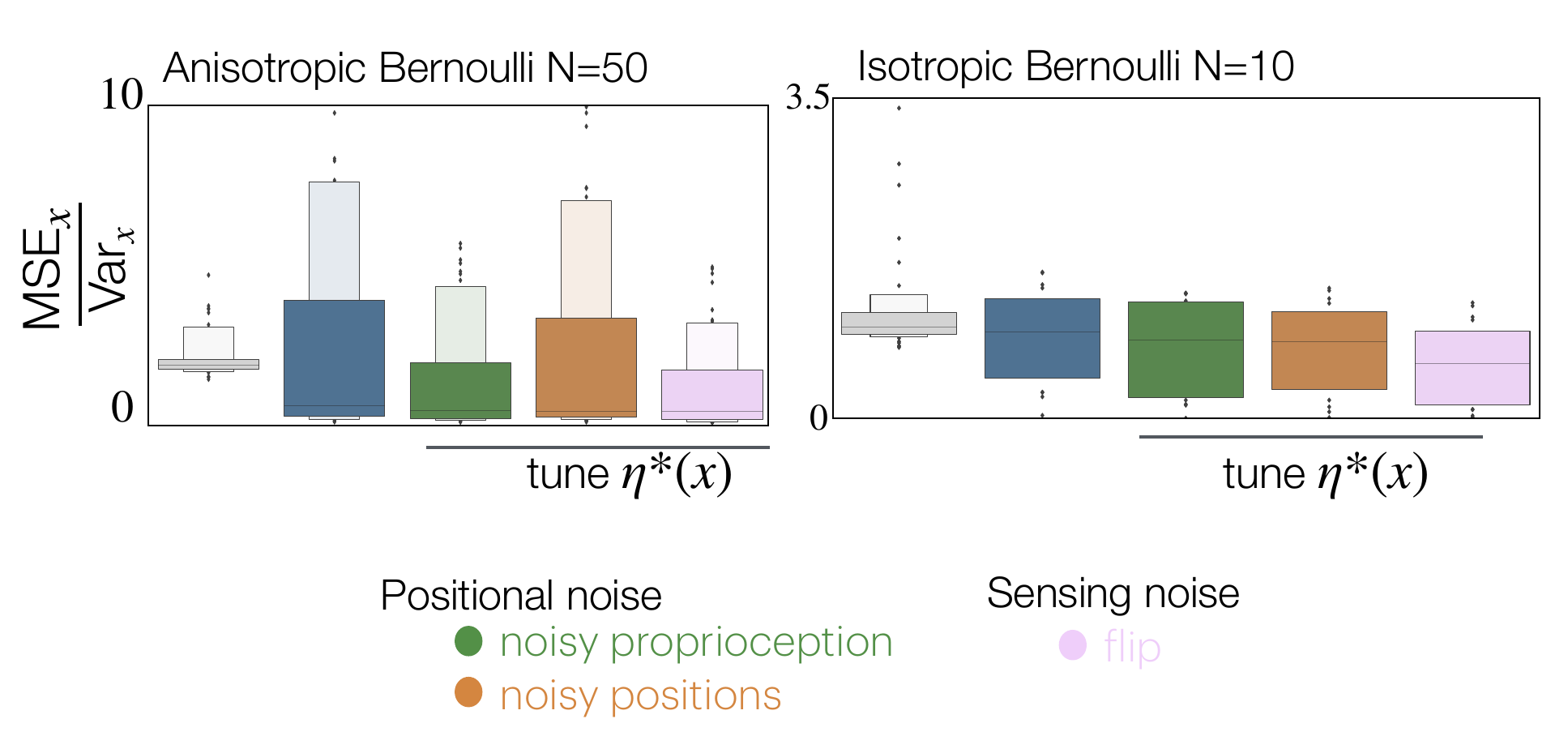} 
\caption{\label{fig:S6}
Aggregate statistics on accuracy of the inference, with different noise sources for anisotropic Bernoulli detections with $N=50$ sensors (left) and isotropic Bernoulli detections with $N=10$ (right). The number of sensors was chosen so as to make the comparison meaningful, i.e.~so that inference with perfect sensing and positional information is inaccurate enough so that it leaves some margin of improvement.
}
\end{figure}

\begin{figure}
\includegraphics[width=0.75\linewidth]{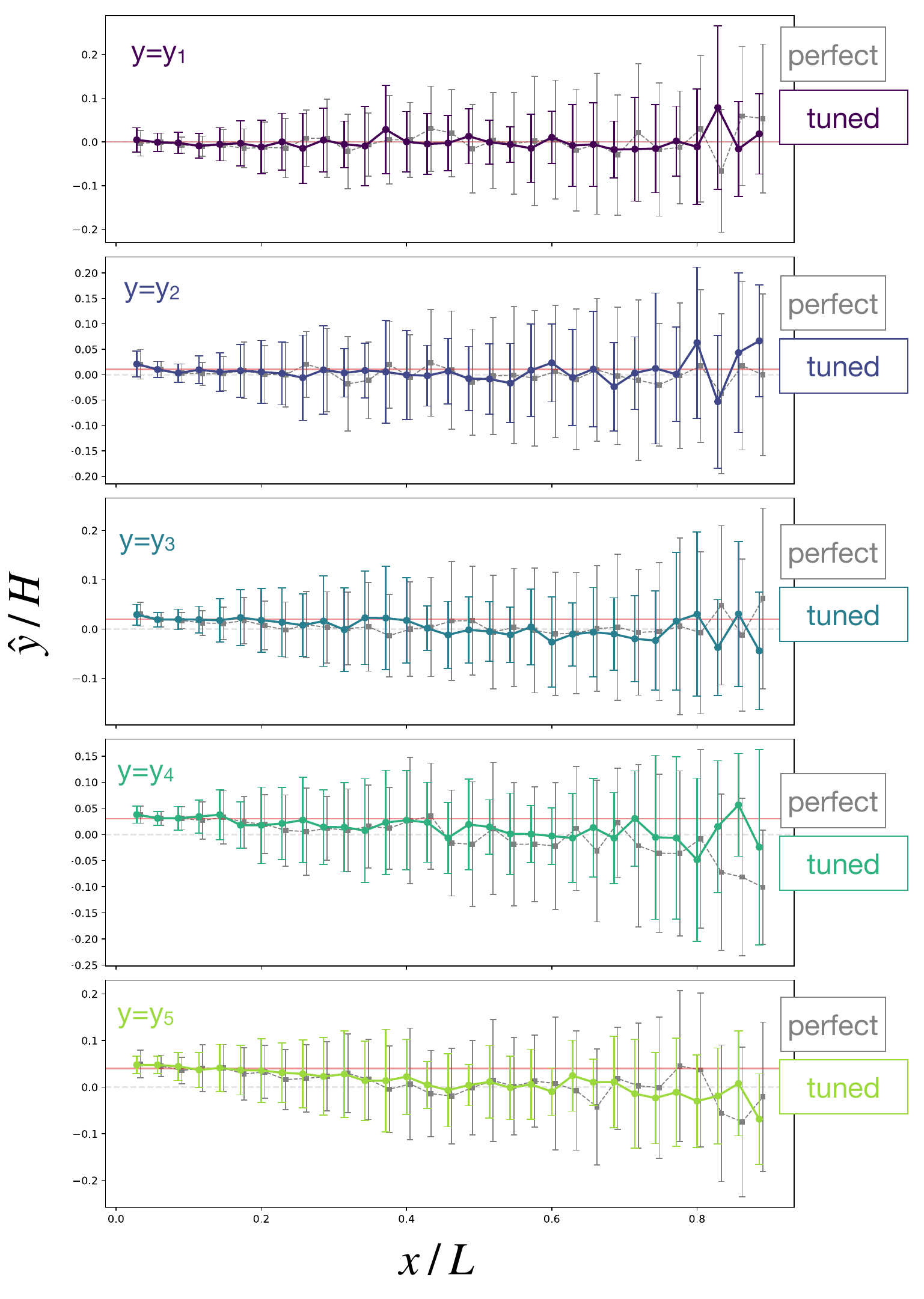} 
\caption{\label{fig:S7}
Predicted $\hat y/H$ for perfect proprioception as a function of $x/L$ (gray) and optimally tuned noisy proprioception (colors), showing that noise in proprioception does not degrade predictions in $y$. Red line corresponds to correct  $y/H$; top to bottom correspond to the five values of $y$ indicated in Fig.~S1, top left (same color code). Here, $L$ is the length of the prior and the width of the prior is $H\approx 0.3 \,L$. 
}
\end{figure}

\begin{figure}
\includegraphics[width=0.75\linewidth]{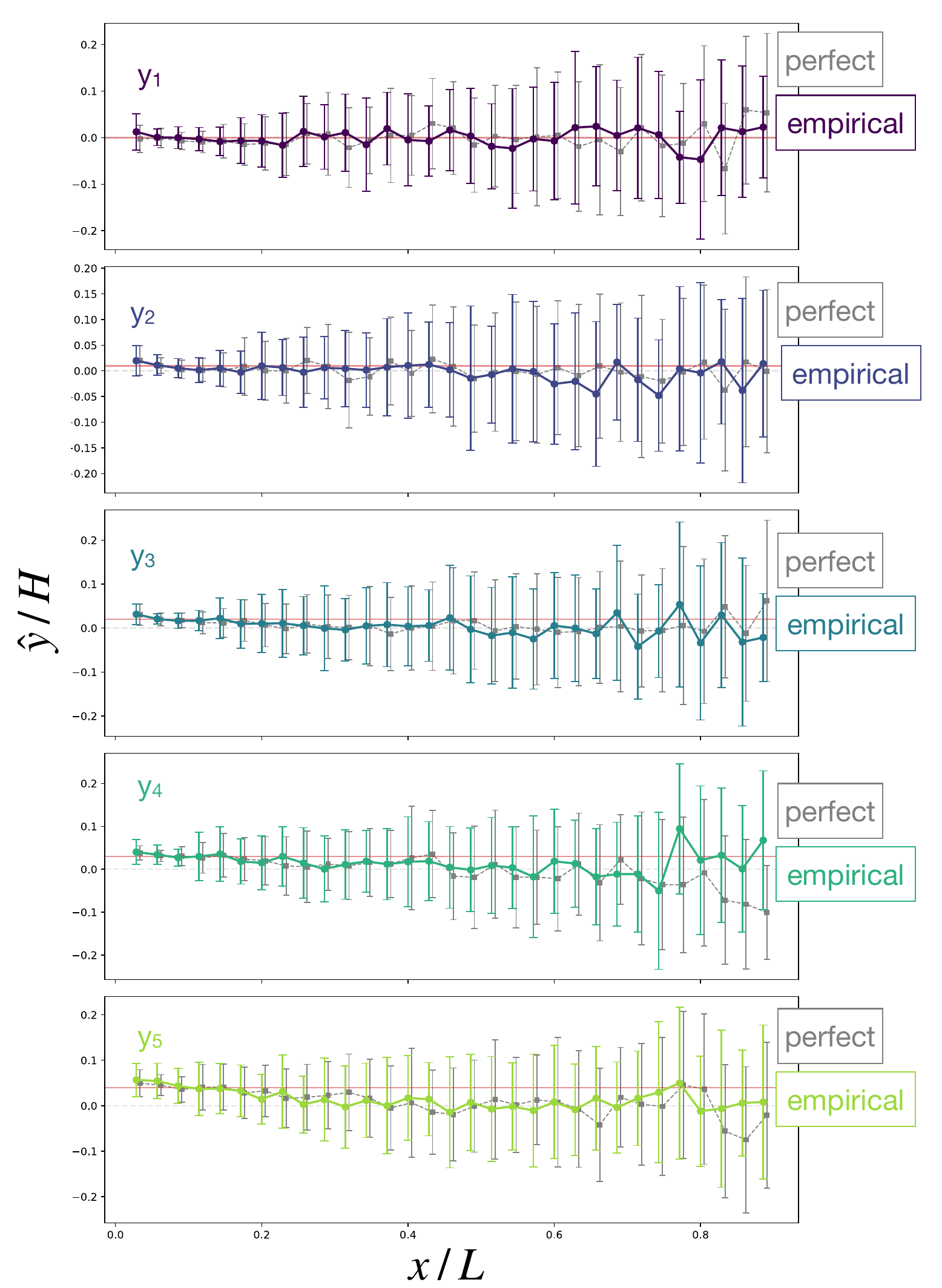} 
\caption{\label{fig:S8}
Same as Fig.~S7, but with noise tuned empirically. 
}
\end{figure}

\begin{figure}
\includegraphics[width=0.75\linewidth]{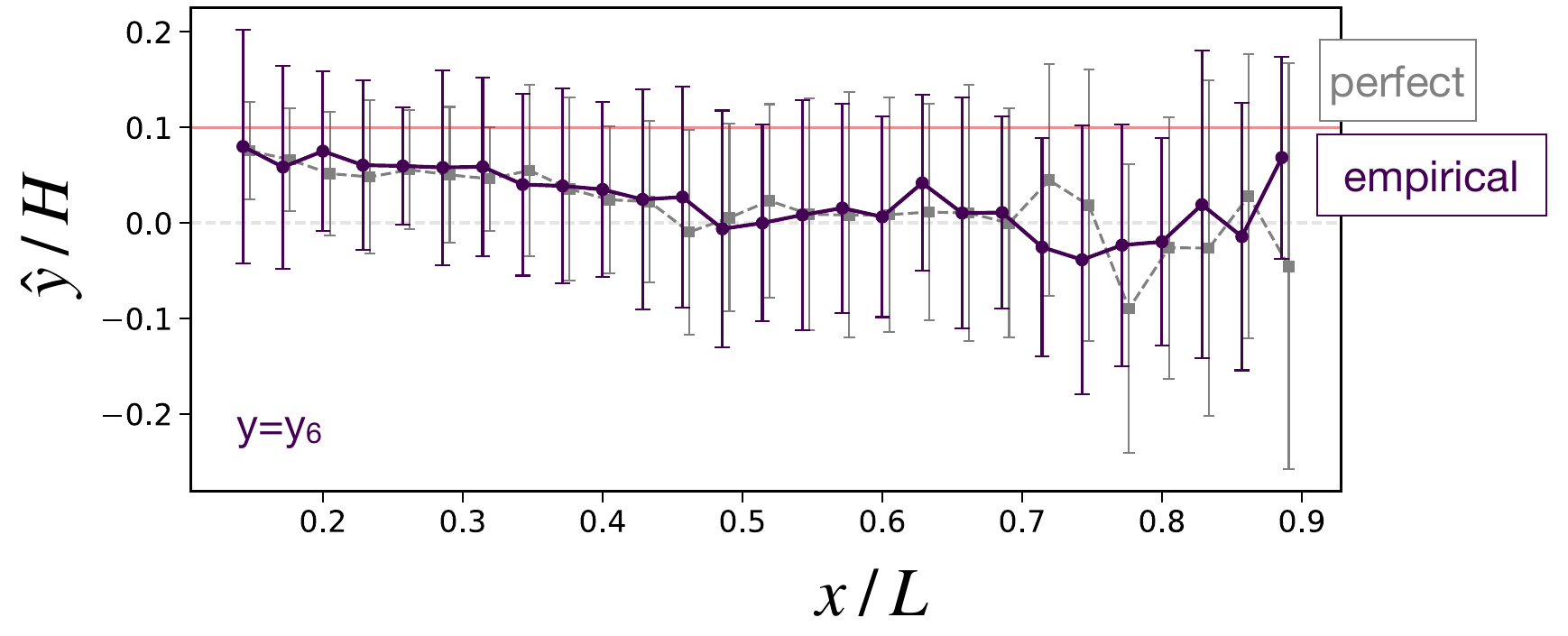} 
\caption{\label{fig:S9}
Same as fig.~S8, for one additional set of points further from the centerline, $y_6/H=0.1$, showing similar qualitative results, with degraded inference near the source. 
}
\end{figure}

\begin{figure}
\includegraphics[width=1\linewidth]{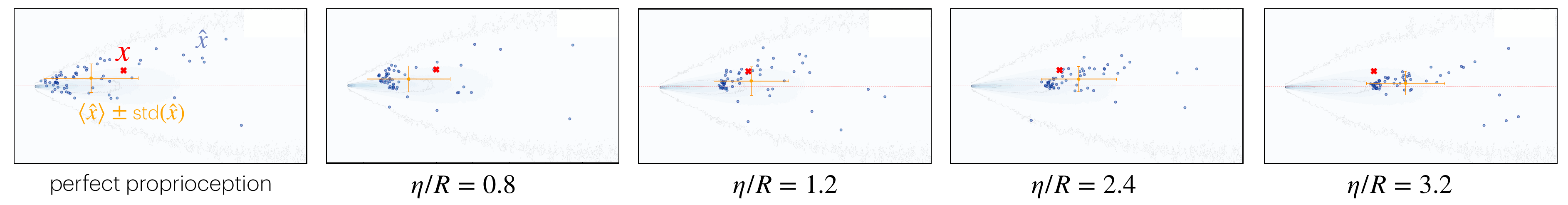} 
\caption{\label{fig:S10}
Inferred locations of the test point using noisy proprioception with $\eta$ set to the value marked under the plot (scatter blue circles); their mean and standard deviation (orange dot and errorbars); real position of the test point (red). From left to right: increasing values of proprioceptive noise move the cloud of inferred positions further away from the source.
}
\end{figure}

\end{document}